\documentclass[12pt]{article}
\usepackage{amsmath}
\usepackage{latexsym}
\usepackage{a4wide}
\usepackage{bbm}
\usepackage{epsfig}
\usepackage{feynmf}
\newcommand{\I}{\text{i}}
\newcommand{\E}{\text{e}}
\newcommand{\tr}{\text{tr}}

\newcommand{\re}[1]{(\ref{#1})}\begin{document}
\newcommand{\sta}[1]{{}^\star\! #1}
\newcommand{\Fk}{F\!k}
\newcommand{\Fqk}{F\!{}^2\!k}
\newcommand{\Bcr}{B_{\text{cr}}}
\unitlength=1mm
\abovedisplayskip17pt plus2pt minus4pt
\abovedisplayshortskip14pt plus2pt minus4pt
\belowdisplayskip17pt plus2pt minus4pt
\belowdisplayshortskip14pt plus2pt minus4pt
\title{QED Effective Action at Finite Temperature: Two-Loop Dominance}
\author{Holger Gies\\
  Institut f\"ur theoretische Physik, Universit\"at T\"ubingen,\\
  72076 T\"ubingen, Germany}
\maketitle
\begin{abstract}
We calculate the two-loop effective action of QED for arbitrary
constant electromagnetic fields at finite temperature $T$ in the limit of
$T$ much smaller than the electron mass. It is shown that in this
regime the two-loop contribution always exceeds the influence of the
one-loop part due to the thermal excitation of the internal photon. As
an application, we study light propagation and photon splitting in the
presence of a magnetic background field at low temperature. We
furthermore discover a thermally induced contribution to pair
production in electric fields.
\end{abstract}

\section{Introduction}

In the low-energy sector of the theory, the effects of quantum
electrodynamics can be summarized in an effective action, the
Heisenberg-Euler action, which enlarges the classical theory of
electrodynamics by non-linear self-interactions of the electromagnetic
field. Technically speaking, this effective action arises from
integrating out the massive (high-energy) degrees of freedom of
electrons and positrons. This program has successfully been carried
out to two-loop order \cite{ritu76-87,ditt85,reut97}. 

The inclusion of finite-temperature effects at the one-loop level has
also been considered in various papers
\cite{ditt79,cox84,loew92,elmf94,shov98,gies99a}, and the real-time
\cite{elmf94} as well as the imaginary-time formalism \cite{gies99a}
finally arrived at congruent results. 

This paper is devoted to an investigation of the thermal QED effective
action at the two-loop level. But contrary to the zero-temperature
case, where the two-loop contribution only represents a 
1\%-correction to the one-loop effective action, we
demonstrate that the thermal two-loop contribution is of a
qualitatively different kind than the thermal one-loop part and
exceeds the latter by far in the low-temperature domain. 

\begin{figure}[h]
\begin{center}
\begin{picture}(125,20)
\put(5,0){
\begin{fmffile}{fmfpic2LoopS}
\begin{fmfgraph*}(120,20)
\fmfleft{i1}
\fmfright{o1}
\fmf{phantom,tension=1}{i1,v1,v2,v3,v4,v5,v6,v7,v8,o1}
\fmffreeze
\fmf{double,left,tension=0.1}{i1,v3}
\fmf{double,left,tension=0.1}{v3,i1}
\fmf{double,left,tension=0.1}{v6,o1}
\fmf{double,left,tension=0.1}{o1,v6}
\fmf{photon}{v6,v7,v8,o1}
\fmfdot{v6,o1}
\end{fmfgraph*}
\end{fmffile}}
\put(0,0){(a)}
\put(75,0){(b)}
\end{picture}
\end{center}

\vspace{0.3cm}

\caption{Diagrammatic representation of the one-loop (a) and two-loop
  (b) contribution to the effective QED action. The fermionic double
  line represents the coupling to all orders to the external
  electromagnetic field.}
\label{figloops}
\end{figure}
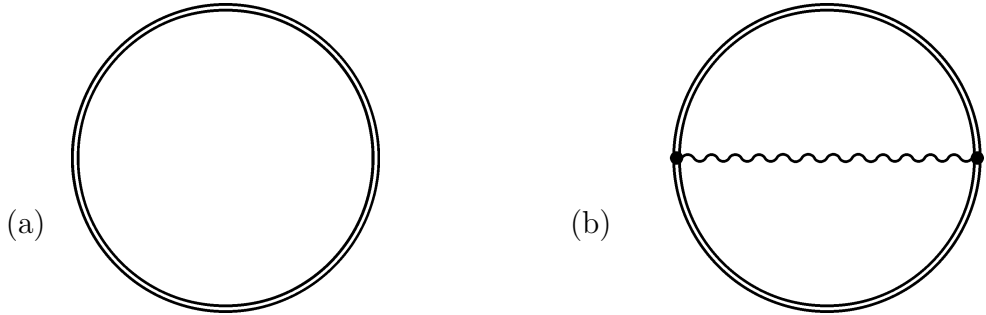

The simple physical reason for this is the following: at one loop, one
takes only the massive electrons and positrons as virtual loop
particles into account (cf. Fig. \ref{figloops}(a)). Due to the mass
gap in the fermion spectrum, a heat bath at temperatures much below
the electron mass $m$ can hardly excite higher fermion states. Hence,
one expects thermal one-loop effects to be suppressed by the electron
mass. In fact, in a low-temperature expansion of the thermal one-loop
effective action \cite{elmf98}, one finds that each term is
accompanied by a factor of $\exp (-m/T)$, exhibiting an exponential
damping for $T\to 0$.

\vspace{1cm}

On the other hand, the two-loop contribution to the thermal effective
action involves a virtual photon within the fermion loop (cf.
Fig.\ref{figloops}(b)). Since the photon is massless, a heat bath of
arbitrarily low temperature can easily excite higher photon states,
implying a comparably strong influence of thermal effects on the
effective action. In Sec. 2, we are able to show that the dominant
contribution to the thermal two-loop effective action in the
low-temperature limit is proportional to $T^4/m^4$. This power-law
behavior always wins out over the exponential damping of the one-loop case,
leading to a {\em two-loop dominance} in the low-temperature domain.
One might ask whether this inversion of the loop hierarchy signals the
failure of perturbation theory for finite-temperature QED. But, of
course, this is not the case, since the inclusion of a virtual photon
does not ``amplify'' the two-loop graph and higher ones. Rather,
calculating the one-loop graph should only be rated as an
inconsistent truncation of the theory, since the one-loop
approximation does not include all species of particles as virtual
ones. Besides, effective field theory techniques indicate that the
three-loop contribution is of the order of $T^8/m^8$ \cite{kong98} for
$T/m \ll 1$, thereby obeying the usual loop hierarchy.

The present paper is organized as follows: In Sec. 2, we present the
calculation of the two-loop effective QED action at finite temperature
employing the imaginary-time formalism and concentrating on the
low-temperature limit. The outcome will be valid for slowly varying
external fields of arbitrary strength. 

Section 3 is devoted to an investigation of light propagation at
finite temperature. While, on the one hand, the well-known result for
the velocity shift $\delta v\sim T^4/m^4$ is rediscovered
\cite{tarr83,bart90,lato95,ditt98}, we are also able to determine
further contributions to the velocity shift arising from a non-trivial
interplay between temperature and an additional magnetic background
field.  

In Sec. 4, we study aspects of thermally induced photon
splitting. Therein, we point out that the thermal two-loop
contribution to the splitting process exceeds the zero-temperature and
one-loop contributions in the low-temperature and weak-field limit,
but is negligible in comparison to other thermally induced
scattering processes. 

Sections 3 and 4 are mainly concerned with the limit of a weak
magnetic background field and low-frequency photons ($\omega\ll m$),
and therefore represent only a first glance at these extensive
subjects. In fact, the quantitative results for this energy regime
describe only tiny effects; a relevance for astrophysical topics such
as pulsar physics has not been identified up to now. However, the
intention of the present work is a more categoric one, namely, to
elucidate the mechanism for a violation of the usual loop hierarchy of
perturbative thermal field theories involving virtual massless
particles. 

In Sec. 5, we calculate the thermal contribution to Schwinger's
famous pair-production formula \cite{schw51} for constant electric
background fields in the low-temperature limit. Here, a thermal
one-loop contribution surprisingly does not exist
\cite{elmf94,gies99a}, since the thermal one-loop effective action is
purely real by construction. Hence, the findings of Sec. 5 prove the
existence of thermally induced pair production -- an effect which has
been searched for for 15 years
\cite{cox84,loew92,hall94,gang95,gang98}. In the low-temperature
limit, we find that the situation of a strong electric field is
dominated by the zero-temperature part (Schwinger formula), while the
thermal contribution can become dominant for a weak electric
field. Unfortunately, the experimentally more interesting
high-temperature limit cannot be covered by our approach. 

One last word of caution: the inclusion of electric background fields
in finite-temperature QED is always plagued with the question of how
violently this collides with assumptions on thermal equilibrium. In
fact, electric fields and thermal equilibrium exclude each other,
thus questioning the physical meaning of the results of
Sec. 5 at least quantitatively. However, it is reasonable to assume
the existence of an at least small window of parameters in the
low-temperature and weak-field domain for which the
thermal-equilibrium calculation represents a good
approximation. Moreover, the knowledge of the effective Lagrangian
including a full dependence on all possible field configurations is
mandatory to derive equations of motion for the fields, even in the
limit of vanishing electric fields. 

\section{Two-Loop Effective Action of QED at Low Temperature}

In the following, we will outline the calculation of the two-loop
effective action, concentrating on the low-temperature limit where a
{\em two-loop dominance} is expected. The calculation is necessarily
very technical, wherefore some details are left for the
appendices.\footnote{The primarily phenomenologically interested
  reader may just take notice of the following conventions
  \re{0a}-\re{0e}, then directly consult Eqs. \re{90}-\re{100}, and
  skip the remainder of the section.} 

But before we get down to business, it is useful to clarify our
notation. From the field strength tensor $F^{\mu\nu}$ and its dual
$\sta{F}_{\kappa\rho}=\frac{1}{2}
\epsilon_{\kappa\rho\mu\nu}F^{\mu\nu}$, we can construct the following
standard gauge and Lorentz invariants:
\begin{eqnarray}
{\cal F}&=&\frac{1}{4} F^{\mu\nu}F_{\mu\nu} = \frac{1}{2} \bigl
( \mathbf{B}^2- \mathbf{E}^2 \bigr) \equiv \frac{1}{2} \bigl( a^2-b^2
\bigr), \nonumber\\ 
{\cal G}&=&\frac{1}{4} F^{\mu\nu} \sta{F}_{\mu\nu} = - \mathbf{E\cdot
  B} \equiv ab,\label{0a}
\end{eqnarray}
where, for reasons of convenience, we also introduced the {\em
  secular} invariants 
\begin{equation}
a=\sqrt{\sqrt{{\cal F}^2+{\cal G}^2}+{\cal F}}, \qquad
b=\sqrt{\sqrt{{\cal F}^2+{\cal G}^2}-{\cal F}}, \label{0b}
\end{equation}
and we assumed without loss of generality that a Lorentz system exists
in which the electric and magnetic field are anti-parallel. In this
particular frame, the secular invariants can be identified with the
field strengths: $a=B\equiv |\mathbf{B}|$, $b=E\equiv |\mathbf{E}|$.

When the physical system involves another vector, say, a momentum
4-vector $k^\mu=(k^0,\mathbf{k})$, we can form another field invariant 
(metric: $g=(-,+,+,+)$): 
\begin{eqnarray}
z_k&:=&(k_\mu F^{\mu\alpha})(k_\nu F^{\nu}{}_\alpha) \nonumber\\
&=& |\mathbf{k}|^2 B^2\sin^2\theta_B +|\mathbf{k}|^2 E^2\sin^2\theta_E
-k^2 E^2 +2 k^0\, \mathbf{E\cdot( k\times B)}, \label{0c}
\end{eqnarray}
where $\theta_B$ ($\theta_E$) denotes the angle between the magnetic
(electric) field and the 3-space vector $\mathbf{k}$. 

In relativistic equilibrium thermodynamics, temperature can be
associated with the invariant norm of a 4-vector $n^\mu$: $n^\mu
n_\mu=-T^2$. On the other hand, $n^\mu$ is related to the
4-velocity vector $u^\mu$ of the heat bath by: $n^\mu=T\,
u^\mu$. E.g., in the heat-bath rest frame, $u^\mu$ takes the form:
$u^\mu=(1,0,0,0)$. Hence, we can introduce one further invariant
(beside the temperature itself):
\begin{equation}
{\cal E}=(u_\mu F^{\mu\alpha})(u_\nu F^{\nu}{}_\alpha). \label{0d}
\end{equation}
E.g., in the heat-bath rest frame, ${\cal E}$ simply reduces to ${\cal
  E}=\mathbf{E}^2$.  Since the effective Lagrangian is a Lorentz
covariant and gauge-invariant quantity, it can only be a function of
the complete set of invariants of the system under consideration.
Hence, we expect a finite-temperature effective QED Lagrangian of the
form:
\begin{equation}
{\cal L}={\cal L}({\cal E},{\cal F},{\cal G},T). \label{0e}
\end{equation}
Equipped with these conventions, we now turn to the calculation.

The two-loop contribution to the effective action/Lagrangian ${\cal
  L}^2$ is generally given by the diagram in Fig. \ref{figloops}(b).
This translates into the following formula in coordinate space
\cite{ditt85}:
\begin{equation}
{\cal L}^2=\frac{e^2}{2} \int d^4x'\, \tr\, \Bigl[ \gamma^\mu\,
G(x,x'|A)\, \gamma^\nu\, G(x',x|A) \Bigr]\,
D_{\mu\nu}(x-x'),\label{01}
\end{equation}
where $G(x,x'|A)$ represents the fermionic Green's function for the
Dirac operator in presence of an external electromagnetic field
$A$. $D_{\mu\nu}$ denotes the photon propagator. Throughout the paper,
we assume the background field to be constant or at least slowly
varying compared to the scale of the Compton wavelength; therefore,
the fermionic Green's function can be written as:
\begin{equation}
G(x,x'|A)= \Phi(x,x') \int \frac{d^4p}{(2\pi)^4}\, \E^{\I p(x-x')}\,
g(p),\label{02}
\end{equation}
where $g(p)$ denotes the Fourier transform of $G(x,x'|A)$ depending
only on the field strength, and $\Phi(x,x')$ is the holonomy carrying
the complete gauge dependence of the Green's function. Inserting
Eq. \re{02} into Eq. \re{01} leads us to the object
$\Phi(x,x')\Phi(x',x)\equiv \Phi(\bigcirc)$, where the right-hand side
represents the holonomy evaluated for a closed path. For a
simply connected manifold such as the Minkowski space,
$\Phi(\bigcirc)=1$; hence, it does not contribute to the
zero-temperature Lagrangian. For a non-simply connected  manifold such
as the finite-temperature coordinate space ($\mathbbm{R}\times S^1$),
$\Phi(\bigcirc)$ can pick up a winding number
\cite{gies99a}. However, in the present case, we restrict our
considerations to a situation with zero density, which implies the
existence of a gauge in which $A_0=0$. Then, $\Phi(\bigcirc)=1$ and the
influence of the holonomy can be discarded. 

This leads us to the representation 
\begin{equation}
{\cal L}^2=\frac{\I}{2} \int \frac{d^4k}{(2\pi)^4} \, D_{\mu\nu}(k)\,
\Pi^{\mu\nu}(k) \label{2}
\end{equation}
for the two-loop Lagrangian, where $D_{\mu\nu}(k)$ denotes the photon
propagator in momentum space, and we introduced the one-loop
polarization tensor in an arbitrary constant external background field:
\begin{equation}
\Pi^{\mu\nu}(k)=-\I e^2 \int \frac{d^4p}{(2\pi)^4}\, \tr\, \bigl
[ \gamma^\mu\, g(p)\, \gamma^\nu\, g(p-k) \bigr]. \label{3}
\end{equation}
So we have finally arrived at the well-known fact that the two-loop
effective action can be obtained from the polarization tensor in an
external field by glueing the external lines together.

The transition to finite-temperature field theory can now be made
within the imaginary-time formalism by replacing the momentum
integration over the zeroth component in Eqs. \re{2} and \re{3} by a
summation over bosonic and fermionic Matsubara frequencies,
respectively. E.g., performing this procedure in Eq. \re{3}
corresponds to thermalizing the fermions in the loop. Now we come to
an important point: confining ourselves to the low-temperature domain
where $T\ll m$, we know from the one-loop calculations \cite{elmf98},
\cite{gies99b} that thermal fermionic effects are suppressed by
factors of $\E^{-m/T}$, indicating that the mass of the fermions
suppresses thermal excitations. Hence, thermalizing the polarization
tensor contributes at most terms of order $\E^{-m/T}$ to the two-loop
Lagrangian for $T\ll m$; these are furthermore accompanied by an
additional factor of the coupling constant $\alpha$ and can therefore
be neglected compared to the one-loop terms. At low temperature, it is
therefore sufficient to thermalize the internal photon only in order
to obtain the leading $T$-dependence of ${\cal L}^{2}$.

Since, in Feynman gauge, the photon propagator reads
\begin{equation}
D_{\mu\nu}(k)=g_{\mu\nu}\, \frac{1}{k^2 -\I\epsilon}, \qquad
k^2=-(k^0)^2 +\mathbf{k}^2, \qquad g=(-,+,+,+), \label{4}
\end{equation}
the introduction of bosonic Matsubara frequencies $(k^0)^2\to
-\omega_n^2=-(2\pi Tn)^2$, $n\in\mathbbm{Z}$, leads us to:\footnote{Of
  course, the present calculation does not necessarily have to be
  performed in the imaginary-time formalism. E.g., instead of
  Eq. \re{4}, we could as well work with the real-time representation
  of the thermal photon propagator. We could even use the  
  one-component formalism only, since we merely consider the photon to
  be thermalized. However, from our viewpoint, the calculations in the
  imaginary-time formalism appear a bit simpler since the momentum
  integrals will remain Gaussian. Of course, this might be just a
  matter of taste.} 
\begin{equation}
{\cal L}^{2+2T}=\frac{\I}{2}\, \I T\sum_{\omega_n} \int
\frac{d^3k}{(2\pi)^3}\, \frac{1}{k^2 -\I\epsilon}\,
\Pi^\mu{}_{\mu}(k). \label{5}
\end{equation}
From now on, we write ${\cal L}^2$ for the zero-temperature two-loop
Lagrangian, ${\cal L}^{2T}$ for the purely thermal part, and ${\cal
  L}^{2+2T}$ for their sum. In Eq. \re{5}, we need the trace of the
polarization tensor in constant but otherwise arbitrary
electromagnetic fields. In the literature, there are various equivalent
representations for $\Pi_{\mu\nu}$. For the present purpose, it is
useful to derive our own one which is based on a calculation of
Urrutia \cite{urru79}. Details are presented in Appendix A. 

Inserting representation \re{10} of the Appendix for $\Pi^\mu{}_\mu$
into Eq. \re{5}, we obtain for the Lagrangian:
\begin{eqnarray}
{\cal L}^{2+2T}\!\!\!\!&=& -\frac{T}{2} \frac{\alpha}{2\pi}
\sum_{\omega_n} \int \frac{d^3k}{(2\pi)^3}
\int\limits_0^\infty\!\frac{ds}{s} 
\!\int\limits_{-1}^1\!\frac{d\nu}{2} \frac{\E^{-\I
      s\phi_0}}{a^2+b^2} \frac{eas\,ebs}{\sin eas \sinh ebs}
  \nonumber\\ 
&&\qquad\quad\left[\frac{z_k}{k^2-\I\epsilon} (\tilde{N}_2
  -\tilde{N}_1) +\bigl( 2N_0(a^2\!+\!b^2) 
  +b^2\tilde{N}_2 +a^2\tilde{N}_1\bigr)\right]
  \Biggl|_{(k^0)^2=-\omega_n^2}\!\!\!\!\!\!\!, \label{11}
\end{eqnarray}
where the $\phi_0$, $N_0$, $\tilde{N}_i$ are functions of the
integration variables $s$ and $\nu$ and of the invariants $a$ and $b$;
only $\phi_0$ depends additionally on $z_k$ as defined in Eq. \re{0c}.
Their explicit form can be looked up in Eqs. \re{111225}, \re{15} and
\re{16}. In order to ensure convergence of the proper-time integrals,
the causal prescription $m^2\to m^2-\I\epsilon$ for the mass term in
$\phi_0$ is understood; this agrees with deforming the $s$-contour
slightly below the real axis.

Now, the aim is to perform the $k$-momentum integration/summation;
note that the $k$-dependence is contained in $\phi_0$, $z_k$ (and
$k^2$, of course). Concentrating on this step, we encounter the
integrals:
\begin{eqnarray}
I_1&=& T\, \sum_{\omega_n} \int \frac{d^3k}{(2\pi)^3}\, \, \E^{-\I s
  \phi_0} \biggl|_{(k^0)^2=-\omega_n^2}, \nonumber\\
I_2&=& T\, \sum_{\omega_n} \int \frac{d^3k}{(2\pi)^3}\,\,
  \frac{z_k}{k^2-\I\epsilon}\,  \, \E^{-\I s \phi_0} 
  \biggl|_{(k^0)^2=-\omega_n^2}, \label{12}
\end{eqnarray}
which allow us to write the Lagrangian \re{11} in terms of
\begin{eqnarray}
{\cal L}^{2+2T}&=&-\frac{\alpha}{4\pi}\int\limits_0^\infty\!\frac{ds}{s} 
\!\int\limits_{-1}^1\!\frac{d\nu}{2}
\frac{eas\,ebs}{(a^2+b^2) \sin eas \sinh ebs}\nonumber\\
&&\qquad\qquad \Bigl( (\tilde{N}_2
  -\tilde{N}_1)\, I_2 +\bigl( 2N_0(a^2\!+\!b^2) 
  +b^2\tilde{N}_2 +a^2\tilde{N}_1\bigr) I_1\,\Bigr). \label{11a}
\end{eqnarray}
Employing Eq. \re{15} for $\phi_0$, we can put down the evaluation of
$I_2$ to the one of $I_1$: 
\begin{eqnarray}
I_2&=& T\,  \sum_{\omega_n} \int
  \frac{d^3k}{(2\pi)^3}\,\,  \frac{z_k}{k^2-\I\epsilon}\, \,
  \E^{-\I m^2s} \E^{-A_z z_k}\, \E^{-A_k k^2}
  \biggl|_{(k^0)^2=-\omega_n^2} \nonumber\\
&=& -\frac{\partial}{\partial A_z}\, \int\limits_{A_k}^\infty dA_k'\,
  \, I_1, \label{17}
\end{eqnarray}
where $A_z$ and $A_k$ again are functions of the integration variables
$s$ and $\nu$ and of the invariants $a$ and $b$, and are defined in
Eq. \re{16}. In view of Eq. \re{17}, it is sufficient to consider the
momentum integration/summation for $I_1$ only:
\begin{equation}
I_1\stackrel{\re{15}}{=}T\, \E^{-\I m^2s}\sum_{\omega_n} \int
  \frac{d^3k}{(2\pi)^3}\,\, \E^{-A_z z_k}\, \E^{-A_k
  k^2}\biggl|_{(k^0)^2=-\omega_n^2}. \label{18}
\end{equation}
At this stage, the {\em finite-temperature coordinate frame} as
introduced in \cite{gies99a} becomes extremely useful, since it
enables us to perform the calculation in terms of the invariants. This
coordinate system is adapted to the situation of electromagnetic
fields at finite temperature in a way that the components of any
tensor-valued function of the field strength can be expressed in terms
of the invariants ${\cal E}$, ${\cal F}$, and ${\cal G}$. Again,
details are presented in the appendix (App. B), from where we take the
final formula for the exponent of Eq. \re{18} (cf. Eq. \re{23}):
\begin{eqnarray}
A_z z_k +A_k k^2 \!\!&=&\!\! \bigl( A_k +(a^2\!-\!b^2\!+\!{\cal E})
A_z\bigr)\! \left(\! k^2\!-\!{\scriptstyle
    \frac{A_z\sqrt{d}}{A_z(2{\cal  F} +{\cal  E}) +A_k}}\, k^0\! 
\right)^2 -\frac{(A_k\!+\!a^2A_z)(A_k\!-\!b^2A_z)}{ A_k
  +(a^2\!-\!b^2\!+\!{\cal  E}) A_z} \, (k^0)^2 \nonumber\\
&&\!\! +\left(\! A_z\frac{a^2b^2}{{\cal E}} +A_k\!\right)\!\left(\!k^3
   +{\scriptstyle \frac{A_z \frac{\sqrt{d}{\cal G}}{{\cal E}}}{A_z
       \frac{{\cal G}^2}{{\cal E}} +A_k}}\, k^1\!\right)
 +\frac{(A_k+a^2A_z)(A_k-b^2A_z)}{ A_k\frac{a^2b^2}{{\cal E}} + A_k}\,
 (k^1)^2, \nonumber\\
&&\label{23T}
\end{eqnarray}
where $k^0,k^1,k^2,k^3$ represent the components of the rotated
momentum vector $k^A=e^A{}_\mu k^\mu$, and $e^A{}_\mu$ denotes the
vierbein which mediates between the given coordinate system and the
finite-temperature coordinate frame (cf. Eq. \re{2.3}). Since the
transformation into the new reference frame is only a rigid rotation
in Minkowski space, no Jacobian arises for the measure of the momentum
integral. Hence, only integrals of Gaussian type are present in
Eq. \re{18}, which can easily be performed to give:
\begin{equation}
I_1=T\,\frac{ \E^{-\I m^2 s}}{(4\pi)^{3/2}}\,\, \frac{1}{\sqrt{p\,
      q_a\, q_b}} \sum_{\omega_n} \E^{-\frac{q_a\, a_b}{p}
      \omega_n^2}, \label{26}
\end{equation}
where it was convenient to introduce the short forms:
\begin{equation}
q_a:=A_k+a^2A_z, \qquad q_b:=A_k-b^2A_z, \qquad p:=A_k+
(a^2\!-\!b^2\!+\!{\cal E})A_z. \label{32}
\end{equation}
The sum in Eq. \re{26} can be rewritten with the aid of a Poisson
resummation of the form:
\begin{equation}
\sum_{n=-\infty}^\infty \exp  \bigl( -\sigma (n-z)^2 \bigr) =
\sum_{n=-\infty}^\infty \sqrt{\frac{\pi}{\sigma}}\,\exp
\left(-\frac{\pi^2}{\sigma}\, n^2 -2\pi\I zn \right). \label{27}
\end{equation}
With $z=0$ and $\sigma=(2\pi T)^2 \frac{q_a q_b}{p}$, we obtain for
Eq. \re{26}:
\begin{equation}
I_1\equiv I_1^{T=0}+I_1^T
=\frac{ \E^{-\I m^2 s}}{16\pi^2}\, \frac{1}{q_a q_b}
 +\frac{ \E^{-\I m^2 s}}{8\pi^2}\, \frac{1}{q_a q_b}\,
\sum_{n=1}^\infty  \exp \left( -\frac{p}{q_a q_b} \,
  \frac{n^2}{4T^2}\right), \label{29}
\end{equation}
where we separated the ($n=0$)-term from the remaining sum in
order to find the ($T=0$)-contribution. The first term in Eq. \re{29}
(($n=0$)-term), namely, is independent of $T$ and ${\cal E}$, while
the second term vanishes in the limit $T\to 0$ exponentially. In
App. C, we check explicitly that the first term of Eq. \re{29} indeed
leads to the (unrenormalized) two-loop Lagrangian for arbitrary
constant electromagnetic fields at zero temperature. E.g., for purely
magnetic fields, the representation of Dittrich and Reuter
\cite{ditt85} is rediscovered. 

For our finite-temperature considerations, we will only keep the
second term of Eq. \re{29}, which we denote by $I_1^T$ in the
following. Concerning the formula for ${\cal L}^{2T}$ in Eq. \re{11a},
$I_1^T$ is already in its final form (it will turn out later that this
term is subdominant in the low-$T$ limit and only $I_2^T$ contains the
important contributions). Hence, let us turn to the evaluation of
$I_2^T$, i.e., the thermal part of Eq. \re{17}; for this, we have to
interpret $I_1^T$ as a function of $A_z$ and $A_k$ (remember: $q_a$,
$q_b$ and $p$ are functions of $A_z$ and $A_k$):
\begin{equation}
I_2^T=-\frac{\partial}{\partial A_z} \int\limits_{A_k}^\infty dA_k'\,
I_1^T(A_k', A_z) =-\frac{\partial}{\partial A_z}
\int\limits_{0}^\infty ds'\, I_1^T(s'+A_k,A_z) =:-\frac{\E^{-\I m^2
    s}}{8\pi^2}\sum_{n=1}^\infty\frac{\partial}{\partial A_z}
\,J(A_z), \label{31} 
\end{equation}
where we defined the auxiliary integral:
\begin{equation}
J(A_z)=\int\limits_{0}^\infty ds'\,\frac{1}{(s'+q_a)(s'+q_b)}
\exp\left( -\frac{s'+p}{(s'+q_a)(s'+q_b)}\, \frac{n^2}{4T^2}
\right). \label{34}
\end{equation}
Upon a substitution of the integration variable,\footnote{Resolving
  for $s'=s'(u)$ leads to a quadratic equation from which the positive
  root has to be taken in order to take care of the integral
  boundaries.} 
\begin{eqnarray}
u&:=&\frac{q_a q_b}{p}\, \frac{s'+p}{(s'+q_a)(s'+q_b)}, \label{34a}\\
\Rightarrow\qquad \frac{ds'}{(s'+q_a)(s'+q_b)} &=& 
-\frac{du}{\sqrt{\frac{q_a^2 a_b^2}{p^2} +\frac{2q_aq_b}{p}
    (2p\!-\!q_a\!-\!q_b) u+(q_a\!-\!q_b)^2 u^2}}, \nonumber
\end{eqnarray}
the auxiliary integral becomes:
\begin{equation}
J(A_z)=\int\limits_0^1 \frac{du}{\sqrt{\frac{q_a^2 a_b^2}{p^2}
    +\frac{2q_aq_b}{p}  (2p\!-\!q_a\!-\!q_b) u+(q_a\!-\!q_b)^2 u^2}}
    \, \exp\left(-\frac{n^2}{4T^2} \frac{p}{q_aq_b}\,
    u\right). \label{36} 
\end{equation}
Now we come to an important point: since we only thermalized the
photons, our effective Lagrangian ${\cal L}^{2T}$ is only valid for
$T\ll m$ anyway. Nevertheless, our formulas also contain information
about the high-temperature domain which we should discard, since it is
incomplete. Regarding Eq. \re{36}, the exponential function causes the
integrand to be extremely small for small values of $T$, except where
$u$ is also small. Hence, the auxiliary integral is mainly determined
by the lower end of the integration interval.

Taking these considerations into account, we expand the square root
for small values of $u$ and then extend the integration interval to
infinity (in fact, maintaining 1 as the upper bound only creates terms
of the order $\exp(-(2nm)/T)$, which are subdominant in the
low-temperature limit). The remaining $u$-integration can then easily
be performed for each order in the $u$-expansion; up to $u^2$, we
obtain:
\begin{equation}
J(A_z)=4\frac{T^2}{n^2} -16 \frac{T^4}{n^4} (2p-q_a-q_b) -64
\frac{T^6}{n^6} \bigl((q_a-q_b)^2-3(2p-q_a-q_b)^2\bigr)+ {\cal
  O}(T^8/n^8). \label{69} 
\end{equation}
Upon differentiation, the $T^2$-dependence drops out, and we get
(cf. Eq. \re{32}):
\begin{equation}
\frac{\partial}{\partial A_z}J(A_z) = -2^5 \frac{T^4}{n^4} ({\cal
  F}+{\cal E}) -2^9 \frac{T^6}{n^6} \bigl( {\cal F}^2+{\cal
  G}^2-3({\cal F}+{\cal E})^2 \bigr) A_z +{\cal
  O}(T^8/n^8). \label{70b}
\end{equation}
In this equation, we indeed discover a power-law dependence on the
temperature, which will directly translate into a power-law dependence
of the two-loop effective action after insertion into Eqs. \re{31} and
\re{11a}. Technically speaking, this arises from the fact that the
omnipresent exponential factor $\exp ( -\frac{n^2}{4T^2}
\frac{p}{q_aq_b}\, u)$, which finally causes exponential damping for 
$T/m\to 0$, becomes equal to 1 after the $u$-integration at the lower
bound at $u=0$. 

At this stage, it is important to observe that the $u$-integration
appears only in $I_2^T$ (via the $A_k'$-integration in Eq. \re{12})
and not in $I_1^T$. Therefore, $I_1^T$ will always contain exponential
damping factors in the limit $T\to 0$. Even the remaining proper-time
integrations do not provide for a mechanism similar to the
$u$-integration, since for large $s$, the mass factor $\exp(-\I m^2
s)$ with the causal prescription $m^2\to m^2-\I\epsilon$ causes the
integrand to vanish, and for small $s$, the combination
$\frac{p}{q_aq_b}$ in the exponent becomes:
\begin{equation}
\frac{p}{q_aq_b} =-\frac{4\I}{1-\nu^2} \frac{1}{s} +{\cal
  O}(s). \label{43a}
\end{equation}
Obviously, inserting Eq. \re{43a} into the exponent leads to an
exponential fall off (bearing in mind that the $s$-contour will run
slightly below the real axis). Similar conclusions can be drawn for
the $\nu$-integration. To summarize these technical considerations, we
conclude that only the term containing $I_2^T$ (thermal part of $I_2$)
in Eq. \re{11a} contributes dominantly to ${\cal L}^{2T}$ in the
low-temperature limit. 

Inserting the first and second term of $\frac{\partial}{\partial
  A_z}J(A_z)$ in Eq. \re{70b} successively into Eq. \re{31} and then
into Eq. \re{11a}, we obtain the dominant terms of order $T^4$ and
$T^6$ of the two-loop effective QED Lagrangian at low temperature;
particularly for the $T^4$-term, different useful representations can
be given:
\begin{eqnarray}
{\cal L}^{2T}\Bigl|_{T^4}\!\!&=&-\frac{\alpha\pi}{90}\, T^4\, ({\cal
  F}+{\cal E}) \int\limits_0^\infty \frac{ds}{s} \int\limits_{-1}^1
  \frac{d\nu}{2}\, \E^{-\I m^2s} \frac{eas\, ebs}{\sin eas \sinh ebs}
  \frac{(\tilde{N}_2- \tilde{N}_1)}{a^2+b^2} \nonumber\\
&=&-\frac{\alpha\pi}{45}\, T^4\, ({\cal F}+{\cal E})
  \int\limits_0^\infty \frac{ds}{s}\, \frac{1}{a^2+b^2}\, \E^{-\I m^2
  s} \nonumber\\
&&\qquad  \left[ ebs \coth ebs \frac{1-eas \cot eas}{\sin^2 eas} 
                +eas \cot eas \frac{1-ebs \coth ebs}{\sinh^2 ebs}
  \right] \label{90}\\
&=&\!\!\!\frac{\pi^2}{45}\, T^4\, ({\cal F}\!+\!{\cal E})\! \left(\!
  \frac{1}{a^2\!+\!b^2} (\partial_a^2\!+\!\partial_b^2)\!\right)\!
  \!\left[\! 
  \frac{1}{8\pi^2}\!\! \int\limits_0^\infty \!\frac{ds}{s^3}\, \E^{-\I
  m^2s} eas \cot eas \, ebs \coth
  ebs\!\right]\!\!. \label{102a} 
\end{eqnarray}
The term proportional to $T^6$ reads:
\begin{equation}
{\cal L}^{2T}\Bigl|_{T^6}\!\! =-\frac{16\alpha\pi^3}{945}\, T^6\,
\bigl( {\cal F}^2\!\!+\!{\cal G}^2\!-\!3({\cal F}\!+\!{\cal E})^2\bigr) 
\!\int\limits_0^\infty \frac{ds}{s} \int\limits_{-1}^1
  \frac{d\nu}{2}\, \frac{\E^{-\I m^2s}}{a^2+b^2}\, 
\frac{eas\, ebs}{\sin eas \sinh ebs}\, (\tilde{N}_2- \tilde{N}_1)\,
A_z, \label{71}
\end{equation}
where $\tilde{N}_i$ and $A_z$ are functions of the integration
variables and the invariants $a$ and $b$ (not of ${\cal E}$), and are
defined in Eqs. \re{111225} and \re{16}. The $\nu$-integration can be
performed analytically, but the extensive result does not provide for
new insights; hence we do not bother to write it down.

These equations represent the central result of the present work;
therefore, a few of their properties should be stressed:

\noindent
1) While we worked explicitly in the low-temperature approximation
$T\ll m$, we put no restrictions on the strength of the
electromagnetic fields. 

\noindent
2) The low-temperature Lagrangians contain arbitrary powers of the
invariants $a$ and $b$ (equivalently ${\cal F}$ and ${\cal G}$), but
the additional invariant at finite temperature ${\cal E}$ only appears
linearly in the $T^4$-term and quadratic in the $T^6$-term. The
small-$T$ expansion thus corresponds to a small-${\cal E}$ expansion.

\noindent
3) The fact that only the integral $I_2^T$ with the prefactor
$(\tilde{N}_2-\tilde{N}_1)$ contributes to the low-temperature
Lagrangian in Eq. \re{11a} implies that only the spatially transversal
modes $\Pi_\|$ and $\Pi_\bot$ of the polarization tensor \re{111224}
play a role in this thermalized virtual two-loop process. The
time-like or longitudinal mode $\Pi_0$ (depending on the character of
$k^\mu$) might become important at higher values of temperature.

\noindent
4) The fact that the invariant ${\cal E}$ always appears in the
combination ${\cal F}+{\cal E}$ ensures a kind of dual invariance of
the Lagrangian. Under the replacement $\mathbf{E}\to \mathbf{B}$ and
$\mathbf{B}\to-\mathbf{E}$, the invariants change into ${\cal F}\to
-{\cal F}$, ${\cal G}\to-{\cal G}$ and ${\cal E}\to{\cal E}+2{\cal
  F}$, so that ${\cal F}+{\cal E}$ remains invariant. 

\noindent
5) The $T^4$-term of ${\cal L}^{2T}$ as exhibited in Eq. \re{102a}
possesses the peculiarity of being derivable from the one-loop
zero-temperature Lagrangian which we marked by square brackets in
Eq. \re{102a} after the derivative terms. This will be elucidated a
bit further in the following section.

For the remainder of this section, we will discuss certain limiting
cases of the two-loop low-temperature Lagrangian. First, let us
concentrate on a weak-field expansion which corresponds to a small-$s$
expansion of the proper-time integral due to the exponential mass
factor. Expanding the integrands for small values of $s$ (except the
mass factor) and integrating over $\nu$ and $s$, leads us to the
dominant terms in the weak-field limit:
\begin{eqnarray}
{\cal L}^{2T}\biggl|_{T^4}&=&\frac{44\alpha^2\pi^2}{2025} 
\frac{T^4}{m^4} ({\cal F}+{\cal E})
-\frac{2^6\cdot37 \alpha^3\pi^3}{3^4\cdot5^2\cdot7} \frac{T^4}{m^4}
\frac{{\cal F}({\cal F}+{\cal E})}{m^4}+{\cal O}(3), \label{73}\\
{\cal L}^{2T}\biggl|_{T^6}&=&\frac{2^{13}\alpha^3\pi^5 }
{3^6\cdot 5\cdot 7^2}
\frac{T^6}{m^6}\bigl(2{\cal F}^2+6{\cal E}{\cal
  F}+3{\cal E}^2-{\cal G}^2\bigr) \frac{1}{m^4}+{\cal
  O}(3), \label{75} 
\end{eqnarray}
where ${\cal O}(3)$ signals that we omitted terms of third order in
the field invariants (sixth order in the field strength). Note that no
linear term in the field invariants to order $T^6$ exists. For the
terms of quadratic order, the $T^6$-term is subdominant for
$T/m\leq0.05$, and amounts up to a 10\%-correction to the $T^4$-term
for $T/m\sim 0.1$. For even larger values of temperature, we expect
the failure of the low-temperature approximation.

Finally, we consider ${\cal L}^{2T}\bigl|_{T^4}$ in the limit of
purely magnetic background fields: $b\to 0$, $a\to B$, ${\cal F}+{\cal
  E}\to \frac{1}{2} B^2$. The $T^4$-term in Eq. \re{90} then reduces
to:
\begin{equation}
{\cal L}^{2T}(B)\biggl|_{T^4}=\frac{\alpha\pi}{90}\,
T^4\int\limits_0^\infty \frac{dz}{z}\, \E^{-\frac{m^2}{eB}z} \left
  [ \frac{1-z\coth z}{\sinh^2 z} +\frac{1}{3} z\coth z\right],
\label{97}
\end{equation}
where we have performed the substitution $eas\to -\I z$ in concordance
with the causal prescription $m^2\to m^2-\I\epsilon$. Incidentally,
the limit of purely electric fields can simply be obtained by
replacing $B\to -\I E$ and multiplying Eq. \re{97} by $(-1)$. 

Introducing the critical field strength $\Bcr:=\frac{m^2}{e}$, we can
evaluate the integral in Eq. \re{97} analytically
\cite{ditt98}\footnote{We take the opportunity to remark that there is
  a misprint in the corresponding integration result in \cite{ditt98};
  the term $(+1/3)$ has to be replaced by $(+1/6)$
  (cf. Eq. \re{101}).} and obtain: 
\begin{eqnarray}
{\cal L}^{2T}(B)\biggl|_{T^4}&=&\frac{\alpha\pi}{90}T^4\biggl[
\Bigl(
{\scriptstyle \frac{\Bcr^2}{2B^2}}\!-\!{\scriptstyle \frac{1}{3}}\Bigr)
\psi (1\!+ \!{\scriptstyle \frac{\Bcr}{2B}})-\frac{2\Bcr}{B}\ln \Gamma 
  ({\scriptstyle \frac{\Bcr}{2B}}) -\frac{3\Bcr^2}{4B^2}  
  \nonumber\\
&&\qquad\qquad\qquad\qquad\!-\frac{\Bcr}{2B}+\frac{\Bcr}{B}\ln 2\pi
  +\!\frac{1}{6}\!+4 \zeta '(-1,{\scriptstyle \frac{\Bcr}{2B}})
  +\frac{B}{3\Bcr}  \biggr], \label{101}
\end{eqnarray}
where $\psi(x)$ denotes the logarithmic derivative of the
$\Gamma$-function, and $\zeta'(s,q)$ is the first derivative of the
Hurwitz $\zeta$-function with respect to its first argument. 

For strong magnetic fields, $B\gg\Bcr$, the last term in square
brackets in Eq. \re{101} dominates the whole expression, and we find a
linear increase of the effective Lagrangian:
\begin{equation}
{\cal L}^{2T}(B\gg\Bcr)\biggr|_{T^4}=\frac{\alpha\pi}{270}\, T^4\,
\frac{eB}{m^2}. \label{100}
\end{equation}
This contribution remains subdominant compared to the one arising from
pure vacuum polarization $\sim B^2 \ln \frac{eB}{m^2}$, which is not
astonishing, since the magnetization of (real) thermalized plasma
particles is bounded: the spins can maximally be completely
aligned. In contrast, the non-linearities of vacuum polarization set
no such upper bound. Quantitatively, the same result was found for the
thermal one-loop contribution \cite{elmf93}.

\section{Light Propagation}

As a first application, we study the propagation of plane light waves
at finite temperature and in a magnetic background. The subject of
light propagation has recently gained renewed interest due to its
accessibility to current experimental facilities \cite{peng98}. 

In the limit of light of low-frequency $\omega\ll m$, the effective
action for slowly varying fields has proved useful for obtaining
velocity shifts, i.e., refractive indices of QED vacua which are
modified by various external perturbations such as fields and
temperature \cite{heyl97,ditt98}. In this limit of low frequencies and
smooth external perturbations, the terms involving derivatives of the
fields in a derivative expansion of the effective action can be
neglected, and the constant-field approximation is appropriate.

The case of light propagation at finite temperature has been
investigated in \cite{gies99b} from a general viewpoint for a class of
Lagrangians depending on the invariants ${\cal E},{\cal F},{\cal G},
T$ in an arbitrary way. Therein, a light cone condition representing a
sum-rule for the polarization modes of the propagating light has been
derived; this has been exploited for a detailed investigation of light
propagation at finite temperature to one-loop order by an insertion of
the thermal one-loop effective Lagrangian of QED. It has been
emphasized that these one-loop studies apply to a domain of
intermediate values of temperature $\sim 0.1 \leq T/m \leq \sim 1$,
where two-loop as well as plasma effects remain subdominant. 

The famous results for the low-temperature velocity shift $\delta
v\sim T^4/m^4$ \cite{tarr83,bart90,lato95} could not have been
rediscovered by this first-principle investigation, because the
thermal two-loop effective action was not at hand. In the present
work, we intend to fill this last gap.

Let us first consider the situation of a thermalized QED vacuum
without an additional background field. In the low-temperature domain,
this vacuum is then characterized by the Lagrangian ${\cal L}=-{\cal
  F}+{\cal L}^{2T}$, where $-{\cal F}$ represents the classical
Maxwell term. Following the lines of \cite{gies99b}, the phase
and group velocity $v$ of a propagating plane wave is then given by:
\begin{equation}
v^2=\frac{1}{1+\frac{2\, \partial_{\cal E}{\cal L}}{(-\partial_{\cal F}
    {\cal L} +\partial_{{\cal E}}{\cal L})}}, \label{2.6}
\end{equation}
where $v=\frac{k^0}{|\mathbf{k}|}$ is constructed from the wave vector
of the propagating light, and it is understood that the partial
derivatives of ${\cal L}$ are evaluated in the zero-field limit.
Inserting Eqs. \re{73} and \re{75} into Eq. \re{2.6}, leads us to:
\begin{equation}
v^2=\frac{1}{1+2\frac{44}{2025} \alpha^2\pi^2
\frac{T^4}{m^4}}\simeq 1-2\frac{44}{2025} \alpha^2\pi^2
\frac{T^4}{m^4}+{\cal O}(T^8/m^8). \label{2.7}
\end{equation}
Note that there is no $T^6$-term, since ${\cal L}^{2T}|_{T^6}$ is at
least quadratic in the field invariants. In Eq. \re{2.7}, we
rediscovered the well-known velocity shifts for light propagation in a
thermal background as found in \cite{tarr83,lato95} via the two-loop
polarization operator and in \cite{bart90,ditt98} via considering vacuum
expectation values of field bilinears in a thermal background. The
here-presented rederivation within the effective action approach from
first principles thus can be viewed as an independent check of
our calculations and of the light cone condition as derived in
\cite{gies99b}. 

But we can go one step further and additionally take a weak external
magnetic field into account; the light cone condition in this case
reads \cite{gies99b}:
\begin{equation}
0=\bigl(\partial_{\cal F}{\cal L}\!-\partial_{{\cal E}}{\cal
  L}\!-{\cal F}\partial^2_{\cal G}{\cal L} \bigr)k^2\! 
+\frac{1}{2} \bigl( \partial^2_{\cal F} \!+\partial^2_{\cal G}\bigr)
  {\cal L}\, z_k
+ 2\partial_{\cal E}{\cal L}\, 
  (ku)^2,\label{2.14}
\end{equation}
where $u^\mu$ denotes the 4-velocity vector of the heat bath and $z_k$
is defined in Eq. \re{0c}. The Lagrangian describing a thermal QED
vacuum with weak magnetic background fields at finite temperature is
given by ${\cal L}=-{\cal F}+{\cal L}^{1}+{\cal L}^{2T}$, where ${\cal
  L}^1$ denotes the one-loop effective Lagrangian at zero temperature.
Up to the second order in the invariants, this famous Heisenberg-Euler
Lagrangian ${\cal L}^1$ is given by:
\begin{equation}
{\cal L}^1=\frac{8}{45} \frac{\alpha^2}{m^4} {\cal F}^2 +
\frac{14}{45} \frac{\alpha^2}{m^4} {\cal G}^2. \label{HE}
\end{equation}
Inserting all the relevant contributions to ${\cal L}$ into the light
cone condition Eq. \re{2.14}, the light velocity to lowest order in
the parameters $T$ and $B$ finally yields:
\begin{equation}
v^2\!=1-\frac{22}{45} \frac{\alpha^2}{m^4} B^2 \sin^2\theta_B -2
\frac{44}{2025} \alpha^2\pi^2 \frac{T^4}{m^4} + \frac{22}{45}
\frac{\alpha^2}{m^4} \left(\!
  \frac{2^5\cdot37}{3^2\cdot5\cdot7\cdot11} \alpha\pi^3
  \frac{T^4}{m^4}\! \right) B^2(1+\sin^2 \theta_B)\!, 
\label{LP7}
\end{equation}
where $\theta_B$ denotes the angle between the propagation direction
and the magnetic field (cf. Eq. \re{0c}). The second and third term
are the well-known velocity shifts for purely magnetic
\cite{adle71,bial70} 
and purely thermal vacua (cf. Eq. \re{2.7}), respectively. The last
term describes a non-trivial interplay between these two vacuum
modifications. The latter can best be elucidated in the various limits
of the angle $\theta_B$; for orthogonal propagation to the magnetic
field $\theta_B=\pi/2$, we get:
\begin{equation}
v^2=1-2\frac{44}{2025} \alpha^2\pi^2 \frac{T^4}{m^4}
  -\frac{22}{45} \frac{\alpha^2}{m^4} B^2 \left
    ( 1-(0.15...)\cdot\frac{T^4}{m^4}\right). \label{LP8}
\end{equation}
For parallel propagation to the magnetic field $\theta_B=0$, we find:
\begin{equation}
v^2=1-2\frac{44}{2025} \alpha^2\pi^2 \frac{T^4}{m^4} \left
  ( 1-(0.96...)\cdot
  \left(\!\frac{eB}{m^2}\!\right)^2\right).\label{LP9}
\end{equation}
Since $T/m$ and $eB/m^2$ are considered to be small in each case, the
corrections to the pure effects in the mixed situation are comparably
small. Note that the mixed thermal and magnetic corrections always
diminish the values for the velocity shift of the pure magnetic or
thermal situations. Let us finally remind the reader that the
here-given velocities hold for low-frequency light ($\omega\ll m$)
only, and represent averages over the two possible polarization modes.
While for the purely thermal case the polarization modes cannot be
distinguished, the situation involving an electromagnetic field
generally leads to birefringence due to the existence of a preferred
direction of the field lines. \bigskip

Let us finally comment on the earlier works \cite{bart90,ditt98}
related to the issue of light propagation in a thermal
background. The philosophy therein was to calculate the velocity
shifts in a purely (weak) electromagnetic background first, and then
take thermal vacuum expectation values of the field
bilinears. Expressing this in formulas, we first recall the expression
for the propagation-direction-averaged light velocity in a weak
electromagnetic background from \cite{ditt98}:
\begin{equation}
v^2=1-\frac{2}{3} (\partial_{{\cal F}}^2 +\partial_{{\cal G}}^2){\cal
  L}\,\, T^{00}, \label{LP10}
\end{equation}
where $T^{00}=\frac{1}{2} (E^2+B^2)$ denotes the 00-component of the
energy-momentum tensor, i.e., energy density of the electromagnetic
field. In the weak-field limit, $(\partial_{{\cal F}}^2
+\partial_{{\cal G}}^2){\cal  L}$ is field independent: $2\frac{22}{45}
\frac{\alpha^2}{m^4}$ (cf. Eq. \re{HE}); therefore, taking thermal
vacuum expectation values of the field quantities in Eq. \re{LP10}
is simply equivalent to replacing $T^{00}$ by $\langle
T^{00}\rangle^T=\frac{\pi^2}{15} T^4$. This then leads to the correct
result as given in Eq. \re{2.7}. 

From the viewpoint of the present work, the correctness of the
approach of \cite{bart90,ditt98} arises from the special form of the
low-temperature two-loop Lagrangian ${\cal L}^{2T}|_{T^4}$ as given in
Eq. \re{102a}. Since 
\begin{equation}
\frac{1}{a^2+b^2}(\partial_a^2+\partial_b^2) =\partial_{\cal
  F}^2+\partial_{\cal G}^2, \label{LP11}
\end{equation}
Eq. \re{102a} can also be written as:
\begin{equation}
\partial_{\cal E} {\cal L}^{2T}\biggr|_{T^4} =\frac{2}{3} \langle
T^{00}\rangle^T\, \frac{1}{2} (\partial_{\cal
  F}^2+\partial_{\cal G}^2){\cal L}^1. \label{LP12}
\end{equation}
Incidentally, Eq. \re{LP12} holds for arbitrary field strength, but, in
this line of argument, it is required for weak fields only. Inserting
Eq. \re{LP12} into the correct light cone condition at finite
temperature, i.e., Eq. \re{2.6}, we obtain to lowest order:
\begin{equation}
v^2\simeq1-2\partial_{\cal E}{\cal L} =1-\frac{2}{3} (\partial_{{\cal
  F}}^2 +\partial_{{\cal G}}^2){\cal  L}\,\,\langle T^{00}\rangle^T,
  \label{LP13} 
\end{equation}
which is equal to the heuristically deduced light cone condition for a
thermal QED vacuum \cite{bart90,ditt98}. 

Note that the combined low-temperature/weak-field effects as given in
Eqs. \re{LP7}-\re{LP9} could not have been found in \cite{ditt98},
since the invariant structure is not completely taken into account in
the heuristic approach. Whether the investigation of the
intermediate-temperature domain to two-loop has been correctly
modeled with the heuristic approach in \cite{ditt98}, cannot be
judged within the present work. Note, however, that the
intermediate-temperature domain is controlled by one-loop effects,
leading to a maximum velocity shift of $-\delta v_{\text{max}}^2
=\frac{\alpha}{3\pi}$ \cite{gies99b}. As has been shown therein, the
{\em two-loop dominance} is lost for $T/m\geq0.058$.

\section{Photon Splitting}

Photon splitting in magnetic fields at zero temperature has been
discussed comprehensively by Adler \cite{adle71}, stressing its
relevance for the photon physics of compact astrophysical objects (see
also \cite{bari95}). For the description of the splitting process for
low-frequency photons with $\omega\ll m$ at weak magnetic fields
$\frac{eB}{m^2}\ll 1$, the use of the one-loop effective Lagrangian
for weak fields is sufficient for obtaining a good estimate of the
absorption coefficient for photon splitting. To be precise, the lowest
order contribution to the splitting process comes from the terms of
third order in the invariants (sixth order in the field strength) of
${\cal L}^1$, i.e., the hexagon graph with one incoming, two outgoing
photons and three couplings to the external magnetic field. Neglecting
dispersion effects, the box graph vanishes because of ${\cal L}^1$
depending on ${\cal F}$ and ${\cal G}$ only, and because of the
Lorentz kinematics of the photons.\footnote{Taking dispersion effects
  into account, the box graph still is only an order $\alpha$
  correction to the hexagon graph.}

The question of thermally induced photon splitting has recently been
investigated by Elmfors and Skagerstam \cite{elmf98} with the aid of
the thermal one-loop effective QED Lagrangian; their studies were
motivated by the fact that a vacuum may be a bad approximation for the
surroundings of some astrophysical compact objects, while a
thermalized environment at zero or finite density might be more
appropriate. It turned out that, at temperatures and magnetic fields at
the scale of the electron mass, the thermal contribution can exceed
the zero-temperature one, but these effects then are superimposed by
Compton scattering of the photons with the plasma. In realistic
situations, the thermally induced process will thus be of subdominant
importance. 

In the following, we intend to complete these results about thermally
induced photon splitting with the dominant low-temperature
contributions stemming from the two-loop process. Hereby, we also
concentrate on the splitting process ($\bot \to \|_1+\|_2$), where a
photon, with its electric field vector orthogonal ($\bot$) to the
plane spanned by the external magnetic field and the propagation
direction, splits into two photons with their electric field vectors
within ($\|$) that plane.\footnote{Note that Adler's definition for
  the $\|,\bot$-mode rely on the direction of the magnetic field
  vector of the photon and thus are opposite to ours.} This is the
only allowed process when dispersion effects are taken into account.

As pointed out in \cite{elmf98}, the box-graph no longer vanishes at
finite temperature, since the Lagrangian now involves an additional
invariant. Hence, the lowest-order contribution to the
photon-splitting matrix element is already produced by the terms of
quadratic order in the invariants in Eqs. \re{73} and \re{75}. 

Without going into details, we recall that the splitting amplitude is
obtained by attaching the external photon legs to the fermion loop,
i.e., differentiating the effective action (which is represented by
the loop) thrice with respect to the fields and then contracting the
result with the field strengths of the involved photons. Hereby, one
has to take into account that the effective Lagrangian now depends on
three field invariants: ${\cal E},{\cal F}$, and ${\cal G}$. The
thermal amplitude arising from the box-graph finally yields:
\begin{equation}
{\cal M}(\bot\to\|_1+\|_2)=2\omega\omega_1\omega_2\, B\sin\theta_B
\,\partial_{{\cal E}{\cal F}} {\cal L}, \label{PS15}
\end{equation}
where $\omega,\omega_1,\omega_2$ denote the frequencies of the
incoming and the two outgoing photons, respectively, and $\theta_B$
again represents the angle between the propagation direction and the
magnetic field. From the splitting amplitude, we obtain the absorption
coefficient $\kappa$ via the formula:
\begin{equation}
\kappa=\frac{1}{32\pi\omega^2} \int\limits_0^\omega d\omega_1
\int\limits_0^\omega d\omega_2\, \delta(\omega-\omega_1-\omega_2)\,\,
{\cal M}^2. \label{PS16}
\end{equation}
Inserting Eq. \re{PS15} for the thermal splitting amplitude into
Eq. \re{PS16} leads us to:
\begin{equation}
\frac{\kappa}{m} =\frac{1}{2^6\cdot3\cdot 5\pi^2}
\left(\frac{eB}{m^2}\right)^2 \sin^2\theta_B 
\left( \frac{\omega}{m}\right)^5 (\partial_{{\cal E}{\cal
      F}}{\cal L})^2 \,m^8. \label{PS20}
\end{equation}
Here, we encounter the typical $(\omega/m)^5$-dependence of the
photon-splitting absorption coefficient for low-frequency photons. The
appearance of the magnetic field to the second power is directly
related to the fact that the box-graph exhibits only one coupling to
the external field. In contrast, Adler's result for the absorption
coefficient at zero temperature arising from the hexagon graph reads
\cite{adle71}:
\begin{equation}
\frac{\kappa^{T=0}}{m} =\frac{13^2}{3^5\cdot 5^3\cdot 7^2}
\frac{\alpha^3}{\pi^2} \left(\frac{eB}{m^2}\right)^6
\sin^6\theta_B\left( \frac{\omega}{m}\right)^5. \label{PS19}
\end{equation}
Here, the three couplings to the external magnetic field produce a
$B^6$-dependence of the absorption coefficient. Therefore, any
finite-temperature contribution will exceed the zero-temperature one
for small enough magnetic fields; but, of course, the absorption
coefficients may then become very tiny. 

In order to obtain the one-loop and two-loop absorption coefficients
for thermally induced photon splitting at low temperature, the
derivatives of the corresponding Lagrangian are required in
Eq. \re{PS20}: 
\begin{eqnarray}
\partial_{{\cal E}{\cal F}}{\cal L}^{1T} &=&\left[ \frac{8\alpha^2}{45}
  \left( \frac{m}{T}\right)^2 +\frac{4\pi\alpha^2}{45} \left
  ( \frac{m}{T}\right)^3 \right] \frac{\E^{-\frac{m}{T}}}{m^4},
  \label{PS23}\\
\partial_{{\cal E}{\cal F}}{\cal L}^{2T} &=&\left
  [ -\frac{2^6\cdot37\alpha^3\pi^3}{3^4\cdot5^2\cdot7} \,
  \left(\frac{T}{m}\right)^4 +\frac{2^{14}
  \alpha^3\pi^5}{3^5\cdot5\cdot7^2} \left( \frac{T}{m}\right)^6
  \right] \frac{1}{m^4}, \label{PS21}
\end{eqnarray}
where we made use of the results of \cite{elmf98} for the
low-temperature/weak-field approximation of the one-loop Lagrangian
${\cal L}^{1T}$, and employed Eqs. \re{73} and \re{75} for the
two-loop one. Obviously, inserting the two-loop terms from Eq.
\re{PS21} into Eq. \re{PS20} leads to a power-law dependence of the
absorption coefficient $\sim T^8/m^8$, while the one-loop terms from
Eq. \re{PS23} imply an exponential mass damping $\exp (-2m/T)$ for
$T\to 0$.

As mentioned above, photons of frequencies below the pair-production
threshold are not only exposed to splitting at finite temperature, but
can also scatter directly with the plasma of electrons and
positrons. Following \cite{elmf98}, the absorption coefficient for the
Compton process is given by
\begin{equation}
\frac{\kappa_{\text{C}}}{m} =\frac{\sigma_{\text{C}}}{m}
\frac{2}{\pi^2} \int\limits_0^\infty dp \frac{p^2}{\E^{\omega_e/T}+1},
\label{PS25}
\end{equation}
where $\omega_e$ denotes the fermion energy $\omega_e=\sqrt{p^2+m^2}$,
and the cross section $\sigma_{\text{C}}$ for unpolarized photons at
$\omega/m\simeq1$ is approximately given by:
\begin{equation}
\sigma_{\text{C}}\simeq\frac{4\pi\alpha^2}{3m^2}. \label{PS26}
\end{equation}
Although $\omega/m\simeq1$ formally represents the maximal limit of
validity of our constant-field approximation for the effective action,
we will continue to consider photons of that frequency in the
following, since, on the one hand, this circumvents a suppression of
the absorption coefficients by the common factor $(\omega/m)^5$, and
on the other hand, it has been shown for the hexagon graph in
\cite{adle71} that the difference between $\omega/m=1$- and
$\omega/m\sim0$-calculations is negligible for weak magnetic fields.

Finally, we have to consider another scattering process which arises
from the presence of a heat bath: photon-photon scattering between the
propagating photon and the black-body radiation of the thermal
background. We estimate the absorption coefficient for this process by
\begin{equation}
\frac{\kappa_{\gamma\gamma}}{m}= \frac{\sigma_{\gamma\gamma}
  n_\gamma}{m}, \label{gg1}
\end{equation}
where $n_\gamma$ denotes the density of photons and is given by:
\begin{equation}
n_\gamma=2 \int\frac{d^3 p}{(2\pi)^3} \frac{1}{\E^{\sqrt{p^2}/T} -1}
=\frac{2\zeta(3)}{\pi^2} \, T^3. \label{gg2}
\end{equation}
Here we encounter the Riemannian $\zeta$-function with
$\zeta(3)\simeq1.202$. The total polarization-averaged cross-section
for photon-photon scattering at low frequencies, as one obtains, e.g.,
from the Heisenberg-Euler Lagrangian \cite{eule36}, reads:
\begin{equation}
\sigma_{\gamma\gamma}=\frac{973}{10125} \frac{\alpha^2}{\pi}
\frac{\alpha^2}{m^2} \left( \frac{\omega_{\text{CM}}}{m}\right)^6,
\label{gg3}
\end{equation}
where $\omega_{\text{CM}}$ denotes the frequency of both photons in
the center-of-mass frame. In order to determine $\omega_{\text{CM}}$,
we first have to find the mean frequency at temperature $T$. Averaging
over the thermal probability distribution for the photons, we find the
mean value $\omega_T=\frac{\pi^4}{30\zeta(3)} T\simeq 2.701 T$.
According to relativistic kinematics, the average value for the
CM-frequency $\omega_{\text{CM}}$ is given by $\omega_{\text{CM}}=
\sqrt{\omega \omega_T/2}\simeq 1.16 \sqrt{T\omega}$, where we averaged
over the propagation direction of the thermal photons. Putting
everything together, we obtain for the absorption coefficient for
photon-photon scattering with the thermal background:
\begin{equation}
\frac{\kappa_{\gamma\gamma}}{m} =\frac{7\cdot 139}{2^5\cdot 3^7\cdot
  5^6} \frac{\pi^9}{\zeta(3)^2} \alpha^4 \left(\frac{T}{m}\right)^6
  \left(\frac{\omega}{m}\right)^3 \simeq 5.21\cdot 10^{-11}
  \left(\frac{T}{m}\right)^6   \left(\frac{\omega}{m}\right)^3
  . \label{gg4}
\end{equation}
Since the average frequency of the heat-bath photons is proportional
to the temperature, this formula becomes invalid for $T\sim m$ and
above, because we employed the low-frequency cross-section in
Eq. \re{gg1}. 

It is already clear from a qualitative viewpoint that there must be a
domain where the two-loop splitting process at least exceeds the
one-loop and the Compton contributions due to the power-law dependence
on the temperature. But since $\kappa^{2T}\sim (T/m)^8$ and
$\kappa_{\gamma\gamma}\sim (T/m)^6$, the two-loop contribution will
eventually be surpassed by the photon-photon scattering for $T\to 0$. 

\begin{figure}
\begin{flushleft}
\begin{picture}(145,70)
\put(0,0){
\epsfig{figure=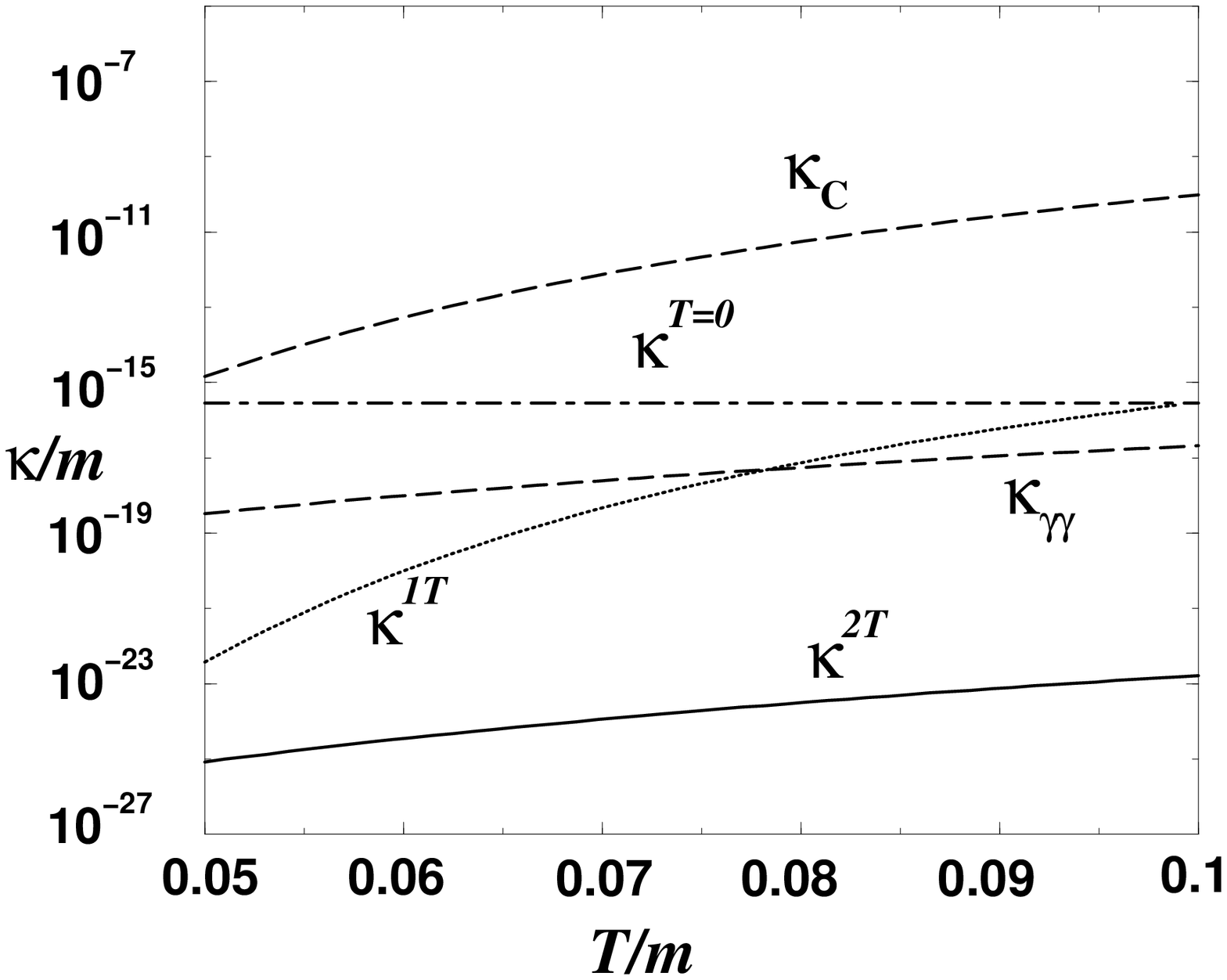,width=7.8cm}
}
\put(0,70){(a): \qquad$\frac{eB}{m^2}=0.2$, $\frac{\omega}{m}=1=\sin
  \theta_B$} 
\put(83,0){
\epsfig{figure=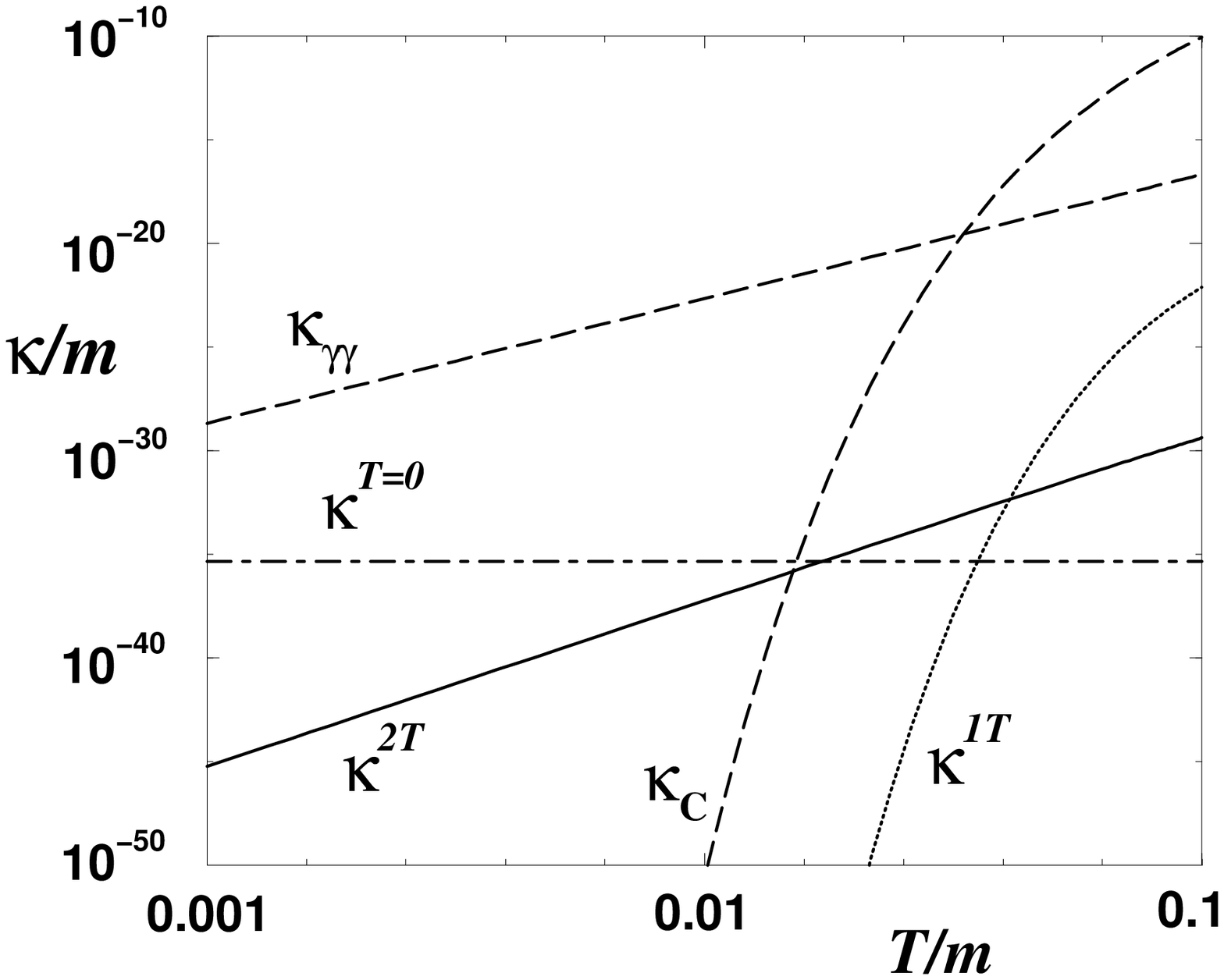,width=7.8cm}
}
\put(85,70){(b):\qquad$\frac{eB}{m^2}=10^{-4}$, $\frac{\omega}{m}=1=\sin
  \theta_B$}
\end{picture}
\end{flushleft}
\caption{Absorption coefficient $\kappa$ in units of the electron mass
  versus temperature $T$ in units of the electron mass. In Fig. (a),
  the various contributions are plotted for parameter values of a
  realistic astrophysical system. In Fig. (b), the parameters are
  chosen in such a way that the two-loop dominance over the one-loop
  and the Compton process is revealed; the photon-photon scattering
  contribution cannot be overtaken in the low-temperure limit.}
\label{figPS1}
\end{figure}

However, quantitative results can only be revealed by numerical
studies.  In fact, as shown in Fig. \ref{figPS1}(a), the two-loop
contribution is completely irrelevant for parameter values which may
be appropriate for a neutron star system and which are close to the
upper bound of validity of our approximation: $\frac{eB}{m^2}=0.2$,
$\omega/m=1$, $\sin \theta_B=1$, and $T/m=0.05\dots0.1$. Even the
one-loop contribution is small compared to the zero-temperature
result; but all are negligible compared to the Compton
process.

Concentrating on the relative strengths of the thermal splitting
processes, the one-loop contribution loses its major role for
$T/m\leq0.041$, where its exponential decrease is surpassed by the
two-loop power law. 

In order to find a domain in which the two-loop splitting wins out
over the zero-temperature process, we have to look at smaller values
of the magnetic field strength; e.g., at values of temperature
$T/m=0.025$, the two-loop process exceeds the zero-temperature one for
$\frac{eB}{m^2} \leq 2.1\cdot 10^{-4}$. Since these are more moderate
field strengths, the absorption coefficient naturally becomes very
small: $\kappa/m\sim 10^{-34}\dots 10^{-33}$. Hence, in order to be
able to measure the splitting rate, the extension of the magnetic
field in which the photon propagates must be comparable to
galactic scales.

Finally, we have plotted the Compton and photon-photon absorption
coefficients, $\kappa_{\text{C}}$ and $\kappa_{\gamma\gamma}$, and the
two-loop coefficient $\kappa^{2T}$ for a 
weak magnetic field $\frac{eB}{m^2}=10^{-4}$ at $T/m=0.001\dots 0.1$
in Fig. \ref{figPS1}(b). Obviously, the Compton process loses its
dominant role for $T/m\leq0.03$; below, the absorption coefficient
is ruled by the photon-photon scattering as long as the temperature
does not become so small that only the zero-temperature amplitude
remains. As is also made visible in Fig. \ref{figPS1}(b), the two-loop
contribution does not exceed the photon-photon process, due to the
weaker temperature dependence of the latter. Hence, we may summarize
that the photon absorption coefficient in the low-temperature domain
is either dominated by the zero-temperature contribution for strong
magnetic fields or by the photon-photon scattering with the thermal
background for weak fields. So the two-loop contribution always
belongs to the top flight but is never ranked first. 

In order to account for realistic astrophysical systems, it is
compulsory to include a finite chemical potential. First estimates can
be found in \cite{elmf98} to one-loop order, where signals have been
found that a finite chemical potential of $\mu\simeq m$ may induce an
increase of the thermal splitting amplitude at low temperatures. In
order to settle this question properly, the present paper shows that
an investigation of these systems should take the two-loop
contributions into account.

Let us conclude this section with the remark that in order to obtain
the sum of the zero-temperature and the thermal contributions to the
photon splitting absorption coefficient, the amplitudes must be summed
up coherently, since the final states of the processes coincide, and
the thermal vacuum with a constant background field does not provide
for a mechanism of decoherence. While the zero-temperature amplitude
as well as the thermal one-loop amplitude are strictly positive, the
$T^4$-term in Eq. \re{PS21} contributes with a negative sign. Hence,
an exceptional curve in the parameter space of $\frac{eB}{m^2}$ and
$T/m$ exists where the thermal two-loop amplitude interferes with the
thermal one-loop and zero-temperature amplitudes destructively so that
photon splitting vanishes. 

\section{Pair Production}

Thermally induced pair production in electric fields has been searched
for at the one-loop level for a long time
\cite{cox84,loew92,hall94,gang95,gang98} with extremely contrary
results. In our opinion, the final concordant judgement in the
real-time formalism \cite{elmf94}, the functional Schr\"odinger
approach \cite{hall94}, as well as the imaginary-time formalism
\cite{gies99a} is that there is no imaginary part in the thermal
contribution to the effective action to one loop, implying the absence
of thermally induced pair production to this order of calculation. As
already mentioned in the introduction, drawing conclusions from an
imaginary part of the thermal effective action to pair production is
not as immediate and straightforward as at zero-temperature, since the
presence of an electric pair-producing field and the thermal
equilibrium assumption which is inherent to our approach contradict
each other.

In the following, we simply assume that on the one hand, the time
scale of pair production is much shorter than the time scale of the
depletion of the electric field so that dynamical back-reactions can
be neglected (this assumption is familiar from the zero-temperature
Schwinger formula). On the other hand, we also assume that the state
of the plasma is appropriately approximated by a thermal equilibrium
although it is exposed to an electric field. Whether the assumption on
thermal equilibrium is justified in concrete experimental situations
such as, e.g., heavy ion collisions, is still under discussion.

Let us now turn to the computation of the imaginary part of the
two-loop thermal effective action for external electric fields. For
this, we concentrate on the $T^4$-contribution as given in
Eq. \re{90}. For purely electric fields, $a\to0$, $b\to E$, ${\cal
  E}+{\cal F}\to\frac{1}{2} E^2$, this reads:
\begin{equation}
{\cal L}^{2T}(E)\biggl|_{T^4}=-\frac{\alpha\pi}{90}\, T^4
\int\limits_0^\infty \frac{dz}{z}\, \E^{-\I \frac{m^2}{eE} z} \left
  [ \frac{1}{3} z \coth z + \frac{1-z\coth z}{\sinh^2 z} \right],
\label{79}
\end{equation}
where we substituted $z=eEs$. For reasons of convenience, it is useful
to abbreviate $\eta:=\frac{eE}{m^2}$, which denotes the dimensionless
ratio between the electric field and the critical field strength
$E_{\text{cr}}:= \frac{m^2}{e}$. Integrating the $1/\sinh^2z$-term by
parts leads us to:
\begin{equation}
{\cal L}^{2T}\!(E)\biggl|_{T^4}\!\!=\!-\frac{\alpha\pi}{90} T^4
\lim_{\epsilon\to 0} \Biggl\{ \frac{1}{2\epsilon^2} +\frac{1}{2} +
\frac{1}{4\eta^2} +\int\limits_\epsilon^\infty dz\, \E^{-\I
  \frac{z}{\eta}} \left(\! \frac{1}{3}-\frac{\I}{\eta z}
  -\frac{1}{z^2} + \frac{1}{2\eta^2} \right) \coth z
\Biggr\}. \label{83} 
\end{equation}
Here, it should be pointed out that the isolated pole in the first
term of the curly brackets does not signal a divergence, but simply
cancels the pole at the lower bound of the integral; the whole
expression is still finite. Our aim is to evaluate the imaginary part
of Eq.  \re{83}; for this, the behavior of the integral at the lower
bound is of no interest. An imaginary part $\text{Im}\, {\cal
  L}^{2T}(E)|_{T^4}$ arises from the poles of the $\coth z$-term on
the imaginary axis at $z=\pm \I n\pi$, $n=1,2,\dots$. 

Decomposing the exponential function into $\cos +\I \sin$, it becomes
obvious that the imaginary parts of the integrand are even functions
in $z$, while the real parts are odd. Thus, extending the integration
interval from $-\infty$ to $\infty$ exactly cancels the real parts and
simply doubles the imaginary parts. We finally get:
\begin{equation}
\text{Im}\, {\cal L}^{2T}(E)\biggl|_{T^4} =-\frac{\alpha\pi}{90}
\frac{T^4}{2\I} \int\limits_{-\infty}^\infty dz\, \E^{-\I
  \frac{z}{\eta}} \left( \frac{1}{3}-\frac{\I}{\eta z} -\frac{1}{z^2}
  + \frac{1}{2\eta^2} \right) \coth z. \label{84a}
\end{equation}
Now we can close the contour in the lower complex half plane, which is
in agreement with the causal prescription $m^2\to m^2-\I
\epsilon$. The value of the integral is then simply given by the sum
of the residues of the $\coth z$-poles at $z=-\I\pi n$,
$n=1,2,\dots$. Hence, we arrive at:
\begin{equation}
\text{Im}\, {\cal L}^{2T}(E)\biggl|_{T^4} =\frac{\alpha\pi^2}{90} \,
T^4 \sum_{n=1}^\infty \E^{-\frac{n\pi}{\eta}} \left( \frac{1}{3}+
  \frac{1}{n\pi\eta} + \frac{1}{n^2\pi^2} + \frac{1}{2\eta^2} \right),
\qquad \eta= \frac{eE}{m^2}, \label{84}
\end{equation}
which represents our final result for the imaginary part of the
thermal effective QED action at low temperature, and should be read
side by side with Schwinger's one-loop result:
\begin{equation}
\text{Im}\, {\cal L}^1(E)=\frac{m^4}{8\pi^3} \, \eta^2
\sum_{n=1}^\infty \frac{\E^{-\frac{n\pi}{\eta}}}{n^2}.\label{84b}
\end{equation}
The sum in Eqs. \re{84} and \re{84b} can be carried out analytically;
but here, it should be sufficient to consider the limiting cases of
weak and strong electric fields.

In the weak-field limit, i.e., for small values of $\eta$, the sum
over $n$ in Eq. \re{84} is dominated by the first term $n=1$.
Furthermore, it is the last term which is the most important one in
parentheses. These considerations then lead us to:
\begin{equation}
\text{Im}\, {\cal L}^{2T}(eE\ll m^2)\simeq \frac{\alpha\pi^2}{180} \,
T^4\, \frac{\E^{-\pi/\eta}}{\eta^2}. \label{85}
\end{equation}
Combining this with the weak-field approximation of Eq. \re{84b}, we
get roughly for the total imaginary part of the effective Lagrangian:
\begin{equation}
\text{Im}\, {\cal L}(eE\ll m^2) = m^4 \E^{-\pi/\eta} \left
  (\! \frac{\eta^2}{8\pi^3} +\frac{\alpha\pi^2}{180} \frac{1}{\eta^2}
  \frac{T^4}{m^4}\! \right) \simeq m^4 \E^{-\pi/\eta} \left(\! 4\cdot
  10^{-3} \eta^2 +4\cdot10^{-4} \frac{T^4/m^4}{\eta^2}\!
  \right). \label{87}
\end{equation}
E.g., for $T/m\simeq0.1$, where the present low-temperature
approximation should still be appropriate, the thermal contribution
can be neglected for $\eta\geq 0.1$; both contributions become roughly
equal for $\eta\simeq 0.056$ (and $T/m=0.1$). For weaker fields and
$T/m\simeq0.1$, the thermal contribution even becomes the dominant
one. 

In the opposite limit, where $\eta\gg 1$, i.e., for strong electric
fields beyond the critical field strength, the 1/3 in parentheses
dominates the expression in Eq. \re{84}, which then gives:
\begin{equation}
\text{Im}\, {\cal L}^{2T}(eE\gg m^2) =\frac{\alpha\pi^2}{270} \, T^4
\sum_{n=1}^\infty \Bigl( \E^{-\pi/\eta}\Bigr)^n
=\frac{\alpha\pi^2}{270}\, T^4 \frac{\E^{-\pi/\eta}}{1-\E^{-\pi/\eta}}
=\frac{\alpha\pi}{270}\, T^4\, \eta +{\cal O}(\eta^0). \label{88}
\end{equation}
Together with the strong-field approximation of the Schwinger formula,
this gives:
\begin{equation}
\text{Im}\, {\cal L}(eE\gg m^2) =m^4 \,\eta\left( \frac{\eta}{48\pi} +
  \frac{\alpha\pi}{270} \frac{T^4}{m^4} \right) \simeq m^4\, \eta
  \left( 6.6\cdot 10^{-3}\eta+8.5\cdot 10^{-5} \frac{T^4}{m^4}
  \right). \label{90z}
\end{equation}
Since Eq. \re{90z} is valid for $\eta\gg 1$ and $T/m\ll 1$, the
low-temperature contribution to $\text{Im}\, {\cal L}(E)$ can be
neglected for strong electric fields. Similarly to the case of strong
magnetic fields, we find that the non-linearities of pure (zero-$T$)
vacuum polarization exceed the polarizability of the thermally induced
real plasma by far in the strong field limit. 

Nevertheless, in the limit of weak electric fields, thermal effects
can increase the pair-production probability $P=1-\exp(-2\text{Im}\,
{\cal L}(E))$ significantly, as was shown in Eq. \re{87}. Of course,
for these values of $\eta$, the total imaginary part is very small due
to the inverse power of $\eta$ in the exponential.

Since we did not consider thermalized fermions, our approach is not
capable of describing high-temperature pair production, which would be
desirable for forthcoming heavy-ion collision experiments. However, as
can be read off from our results for light propagation and photon
splitting, extrapolating the power-law behavior to higher
temperature scales of $T\sim m$ or even $T/m\gg 1$ overestimates a
possible two-loop contribution by far, since, for these values of
temperature, the one-loop contribution can be expected to be the
dominant one. The latter increases at most logarithmically with $T$. 

Therefore, it is reasonable to assume that the pair-production
probability also increases at most logarithmically with $T$. In view
of these considerations, a power-law growth as suggested in
\cite{loew92,gang95,gang98} does not appear plausible. Of course, in
order to decide this question, the two-loop calculation has to be
carried out for arbitrary values of temperature. 

\section{Discussion}

In the present work, we calculated the thermal two-loop contribution
to the effective QED action for arbitrary constant electromagnetic
fields in the low-temperature limit, $T/m\ll 1$. Contrary to the usual
loop hierarchy in a perturbation theory with small coupling, the
thermal two-loop part is found to be dominating over the thermal
one-loop part in the low-temperature limit, since the former exhibits
a power-law behavior in $T/m$, while the latter is exponentially
suppressed by factors of $\exp(-m/T)$. The physical reason behind this
is that the one-loop approximation does not involve virtual photons,
which, due to their being massless, can be more easily excited at
low temperatures than massive fermions; thus, the one-loop
approximation should be rated as an inconsistent truncation of
finite-temperature QED for $T$ much below the electron mass. 

The power-law dependence of the thermal effective action to two loop
starting with $T^4/m^4$ implies a {\em two-loop dominance} in the 
low-energy domain of thermal QED, which holds roughly up to
$T/m\simeq0.05$. 

For the subject of light propagation at finite temperature, this
two-loop dominance has been known for some time from studies of the
polarization tensor \cite{tarr83,lato95}. Moreover, for the subject of
QED in a Casimir vacuum like the parallel-plate configuration, the
two-loop dominance is very natural and well known, since the fermions
are not considered to be subject to the periodic boundary
conditions anyway. This gives rise to a non-trivial check of our results,
since Casimir and finite-temperature calculations highly resemble each
other. Replacing, as usual, $T$ by $1/(2a)$ in Eq. \re{85} for the
weak-field limit of the imaginary part of the effective Lagrangian,
where $a$ denotes the separation of the Casimir plates, we obtain:
\begin{equation}
\text{Im}\, {\cal L}^{2a}(E)\biggl|_{a^{-4}}
=\frac{\pi e^2}{2^8\cdot 45}
\frac{1}{a^4} \left( \frac{m^2}{eE}\right)^2 \E^{-\pi m^2/eE},
\label{99}
\end{equation}
which agrees precisely with the findings of \cite{roba87} for the
Casimir corrections to the Schwinger formula\footnote{Actually,
  Eq. \re{99} agrees with the findings of \cite{roba87} except for a
  global sign; however, as was pointed out by one of the authors in a
  footnote of \cite{scha90}, the expression in \cite{roba87} is wrong
  by a minus sign, which saves the day.}.

In order to illustrate the two-loop dominance, we studied light
propagation and photon splitting in a weak magnetic background at
low temperature. Since we are dealing with the two-loop level, the
here-considered effects are naturally very tiny and a significant
influence on, e.g., photon physics near astrophysical compact objects
appears not very probable. One should rather take a closer look at
photon physics on large galatic scales. 

Furthermore, we calculated the imaginary part of the thermal two-loop
effective action for electric background fields at low temperature.
Under mild assumptions, this result can be related to a thermally
induced production probability of electron-positron pairs. Especially
in the weak-field limit, the thermal contribution has a significant
influence on the production rate. Since no thermal one-loop imaginary
part exists, any finite two-loop result automatically dominates at
any temperature scale.

For the subjects of light propagation and photon splitting, the loop
hierarchy is restored above $T/m\simeq 0.05$. Already at this
comparably low value of temperature, the thermal excitation of the
fermions begins to compete with that of the virtual photon. Hence,
a calculation of the two-loop thermal Lagrangian at intermediate or
high temperatures would appear as an imposition, were it not for the
high-temperature pair-production probability which is beyond the range
of the one-loop approximation and of great interest for, e.g.,
heavy-ion collisions.  

\section*{Appendix}

\renewcommand{\thesection}{\mbox{\Alph{section}}}
\renewcommand{\theequation}{\mbox{\Alph{section}.\arabic{equation}}}
\setcounter{section}{0}
\setcounter{equation}{0}

\section{One-loop Polarization Tensor}

While the polarization tensor in an external magnetic field has been
considered by many authors (a comprehensive study can, e.g., be found
in \cite{tsai75}), a generalization to arbitrary constant
electromagnetic fields in a straightforward manner is associated with
a substantial increase in calculational difficulties. The problem was
first solved by Batalin and Shabad \cite{bata71}; their extensive
result was later brought into a practical form by Artimovich
\cite{arti90}. In the following, we will briefly sketch a simpler
derivation of the polarization tensor in arbitrary constant
electromagnetic fields; our approach is based on the findings of
Urrutia \cite{urru79}, who solved the problem for the special case of
parallel electric and magnetic fields.

\vspace{7mm}

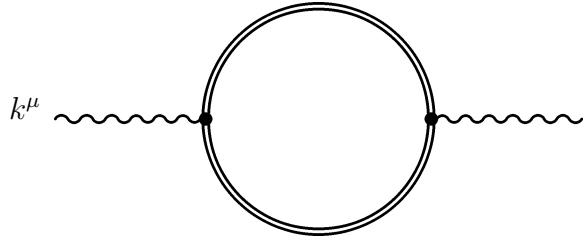
\begin{figure}[h]
\begin{center}
\begin{picture}(125,20)
\put(27,0){
\begin{fmffile}{fmfpicPolTens}
\begin{fmfgraph*}(70,20)
\fmfleft{i1}
\fmfright{o1}
\fmf{phantom,tension=1}{i1,v1,v2,v3,v4,v5,v6,o1}
\fmffreeze
\fmf{photon}{i1,v1,v2}
\fmf{double,left,tension=0.1}{v2,v5}
\fmf{double,left,tension=0.1}{v5,v2}
\fmf{photon}{v5,v6,o1}
\fmfdot{v2,v5}
\end{fmfgraph*}
\end{fmffile}}
\put(25,10){$k^\mu$}
\end{picture}
\end{center}

\vspace{0.3cm}

\caption{Diagrammatic representation of the one-loop polarization
  tensor. The fermionic double line represents the coupling to all
  orders to the external electromagnetic field.}
\label{figpol}
\end{figure}

Assume that the $(-E)$-field and the $B$ field point along the
3-axis. 4-vectors like the external momentum (cf. Fig. \ref{figpol})
can then be decomposed into 
\begin{equation}
k^\mu=k^\mu_\|+k^\mu_\bot, \qquad k^\mu_\|=(k^0,0,0,k^3), \qquad
k^\mu_\bot= (0,k^1,k^2,0). \label{11125}
\end{equation}
In the same manner, tensors can be decomposed, e.g.,
$g^{\mu\nu}=g^{\mu\nu}_\|+ g^{\mu\nu}_\bot$. With respect to each
subspace, we easily find the unique orthogonal vector to a given one:
\begin{equation}
\tilde{k}^\mu_\|=(k^3,0,0,k^0),\qquad\qquad
\tilde{k}^\mu_\bot=(0,k^2,-k^1,0). \label{11126}
\end{equation}
Following Urrutia \cite{urru79}, the polarization tensor for the
special field configuration can be written as:
\begin{eqnarray}
\Pi^{\mu\nu}(k|A})\!=\! \frac{\alpha}{2\pi}\!\!
\int\limits_0^\infty \frac{ds}{s}\!\int\limits_{-1}^1\!
\frac{d\nu}{2}\Biggl\{{&&\!\!\!\!\!\!\! \E^{-\I s\phi_0} \frac{z z'}{\sin
  z \sinh z'} 
\biggl[ \bigl(g^{\mu\nu} k^2 -k^\mu k^\nu \bigr)N_0
+\bigl(g^{\mu\nu}_\| k^2_\| -k^\mu_\| k^\nu_\| \bigr)N_1 \nonumber\\
&&+\bigl(g^{\mu\nu}_\bot k^2_\bot -k^\mu_\bot k^\nu_\bot \bigr)N_2
-\bigl( \tilde{k}^\mu_\bot \tilde{k}^\nu_\| +\tilde{k}^\mu_\|
\tilde{k}^\nu_\bot \bigr) N_3 \biggr]+\text{c.t.}\Biggr\}. \label{11127}
\end{eqnarray}
The electric and magnetic field strengths $E,B$ are contained in the
variables $z:=eBs$ and $z':=eEs$. The exponent $\phi_0$ is given
by\footnote{This formula has been misprinted in Ref. \cite{urru79}.}
\begin{equation}
\phi_0:=m^2+\frac{k^2_\|}{2} \frac{\cosh z'-\cosh \nu z'}{z' \sinh z'}
+\frac{k^2_\bot}{2} \frac{\cos \nu z -\cos z}{z \sin z}. \label{11128}
\end{equation}
The functions $N_i$ read:\footnote{$N_3$ differs from Urrutia's
  findings by a minus sign, since he considers {\em parallel} $E$- and
  $B$-fields.}
\begin{eqnarray}
N_0&=&\cosh \nu z'\, \cos \nu z-\sinh \nu z'\, \sin \nu z\, \cot z\, \coth z',
\nonumber\\
N_1&=&2\cos z \frac{\cosh z' -\cosh \nu z'}{\sinh^2 z'} -N_0\qquad\quad
=: \tilde{N}_1 -N_0, \nonumber\\ 
N_2&=&2\cosh z' \frac{\cos \nu z -\cos z}{\sin^2 z} -N_0\qquad\qquad =:
\tilde{N}_2 -N_0, \label{11129}\\
N_3&=&-\frac{1-\cos z\, \cos \nu z}{\sin z}\frac{\cosh \nu z'\, \cosh z'
  -1}{\sinh z'} +\sin \nu z\, \sinh \nu z', \nonumber
\end{eqnarray}
where we have incidentally defined the functions $\tilde{N}_{1,2}$ for
later use. The determination of the contact term corresponds to a charge
and field strength renormalization and yields:
\begin{equation}
\text{c.t.}=-\E^{-\I m^2 s}(1-\nu^2) \bigl(g^{\mu\nu}k^2-k^\mu
k^\nu\bigr). \label{111210}
\end{equation}

Now, one can show \cite{gies99c} that the Lorentz invariant form of
the polarization tensor for arbitrary constant electromagnetic fields
can be completely reconstructed from the special form given above for
anti-parallel electric and magnetic fields. This is achieved by,
first, a one-to-one mapping between Urrutia's scalar variables
($k_\|^2,k_\bot^2,E,B$) and a set of invariants which reduce to
Urrutia's variables in the special system:
\begin{eqnarray}
a&\to&B, \qquad\qquad\qquad\qquad\quad b\,\,\to-E, \nonumber\\
z_k&\to&-E^2k^2_\| +B^2k^2_\bot, \qquad\quad
k^2\to k^2_\|+k^2_\bot. \label{111212}
\end{eqnarray}
The inverse map is obtained by a simple calculation; the non-trivial
relations are:
\begin{equation}
k^2_\|\to\frac{a^2k^2-z_k}{a^2+b^2}, \qquad\qquad k^2_\bot\to \frac{b^2k^2
  +z_k}{a^2+b^2}. \label{111213}
\end{equation}
Secondly, the reconstruction requires a one-to-one mapping between
Urrutia's tensor structures in Eq. \re{11127} and Lorentz covariant
tensors which reduce to Urrutia's in the special system. For this, we
need to introduce the following definitions. First, we employ a set of
four linearly independent 4-vectors:
\begin{equation}
k^\mu,\quad \Fk^\mu\equiv F^{\mu\alpha}k_\alpha,\quad
\Fqk^\mu\equiv F^{\mu\alpha}F_{\alpha\beta}k^\beta,\quad
\text{and}\quad \sta{\Fk}^\mu\equiv
\sta{F}^{\mu\alpha}k_\alpha. \label{111214}
\end{equation}
From these, we construct the 4-vectors:
\begin{eqnarray}
v^\mu_\|&:=&\frac{1}{a^2+b^2}\Bigl(a\,\sta{\Fk}^\mu -b\, \Fk^\mu
\Bigr)\quad \to \quad \tilde{k}^\mu_\|,\nonumber\\ 
v^\mu_\bot&:=&\frac{1}{a^2+b^2}\Bigl(b\,\sta{\Fk}^\mu +a\, \Fk^\mu
\Bigr)\quad \to \quad \tilde{k}^\mu_\bot,\label{111219} 
\end{eqnarray}
where the subscripts $\|$ and $\bot$ are to remind us of the meaning of
$v_\|$ and $v_\bot$ in the special Lorentz system (longitudinal and
transversal part of $\tilde{k}$). Incidentally, they obey the
relations (cf. Eq. \re{111213}):
\begin{equation}
v^2_\|=v^\mu_\|v_{\|\mu}=-\frac{a^2k^2-z_k}{a^2+b^2},\quad
v^2_\bot=v^\mu_\bot v_{\bot\mu}=\frac{b^2k^2+z_k}{a^2+b^2},\quad
v^\mu_\|\,v_{\bot\mu}=0. \label{111220}
\end{equation}
Finally introducing the projectors
\begin{eqnarray}
P^{\mu\nu}_0&:=& \frac{1}{k^2\left[2{\cal F}\frac{z_k}{k^2}
  +{\cal G}^2-\left(\frac{z_k}{k^2}\right)^2\right]} \left(
  F^2k^\mu+\frac{z_k}{k^2}\,
  k^\mu\right)\left(F^2k^\nu+\frac{z_k}{k^2}\,
  k^\nu\right),\nonumber\\
P^{\mu\nu}_\|&:=&\frac{v^\mu_\| v^\nu_\|}{v^2_\|},\qquad\qquad\qquad
  P^{\mu\nu}_\bot:=\frac{v^\mu_\bot v^\nu_\bot}{v^2_\bot}, 
  \label{111222}
\end{eqnarray}
which satisfy the usual projector identities,
$P_{0,\|,\bot}^2=P_{0,\|,\bot}$,
$P_{0,\|,\bot}{}^\mu{}_\mu=1$, we can establish the one-to-one
mapping:
\begin{eqnarray}
 -v^\mu_\|\,v^\nu_\|&\to& \bigl(g^{\mu\nu}_\| k^2_\| -k^\mu_\|
 k^\nu_\| \bigr),\nonumber\\ 
v^\mu_\bot\, v^\nu_\bot&\to& \bigl(g^{\mu\nu}_\bot k^2_\bot
-k^\mu_\bot k^\nu_\bot \bigr),\nonumber\\ 
Q^{\mu\nu}:= v^\mu_\bot\, v^\nu_\| +v^\mu_\| \, v^\nu_\bot&\to&\bigl
 ( \tilde{k}^\mu_\bot \tilde{k}^\nu_\| +\tilde{k}^\mu_\|
 \tilde{k}^\nu_\bot \bigr), \label{111221}\\
 k^2\Bigl[ P^{\mu\nu}_0
  +P^{\mu\nu}_\|+P^{\mu\nu}_\bot\Bigr] &\to&\bigl(g^{\mu\nu} k^2
  -k^\mu k^\nu \bigr). \nonumber 
\end{eqnarray}
In the third line, we have defined the object $Q^{\mu\nu}$, which is
neither a projector nor orthogonal to the $P^{\mu\nu}_{\|,\bot}$'s but
orthogonal to $P^{\mu\nu}_0$. 

We are finally in a position to transform the polarization tensor
for the parallel field configuration into its generalized form for 
arbitrary constant electromagnetic fields:
\begin{equation}
\Pi^{\mu\nu}(k|A)=\Pi_0\, P_0^{\mu\nu} +\Pi_\|\, P_\|^{\mu\nu}
+\Pi_\bot\, P_\bot^{\mu\nu} +\Theta\, Q^{\mu\nu}, \label{111223}
\end{equation}
where $\Pi_{0,\|,\bot}$ and $\Theta$ are functions of the invariants
and read:
\begin{equation}
\left\{ \begin{array}{c} \Pi_0 \\ \Pi_\| \\ \Pi_\bot \\ \Theta
  \end{array} \right\} =\frac{\alpha}{2\pi}\!\!
\int\limits_0^\infty \frac{ds}{s}\!\int\limits_{-1}^1\!
\frac{d\nu}{2}\Biggl[ \E^{-\I s\phi_0}\frac{eas\,ebs}{\sin eas \sinh
  ebs} \left\{ \begin{array}{c}  k^2N_0 \\  N_0v^2_\bot -\tilde{N}_1
    v^2_\| \\ \tilde{N}_2 v^2_\bot-N_0v^2_\| \\ -N_3 \end{array}
\right\} + \text{c.t.} \Biggr]. \label{111224}
\end{equation}
Substituting the invariants into Eqs. \re{11128} and \re{11129}, the
functions $N_i$ and $\phi_0$ yield:
\begin{eqnarray} 
\phi_0&=&m^2-\frac{v^2_\|}{2} \frac{\cosh ebs-\cosh \nu ebs}{ebs \sinh
  ebs} +\frac{v^2_\bot}{2} \frac{\cos \nu eas -\cos eas}{eas \sin eas},
  \nonumber\\ 
N_0&=&\cosh \nu ebs\, \cos \nu eas-\sinh \nu ebs\, \sin \nu eas\, \cot
  eas\, \coth ebs, \nonumber\\
\tilde{N}_1&=&2\cos eas \frac{\cosh ebs -\cosh \nu ebs}{\sinh^2 ebs},
  \nonumber\\  
\tilde{N}_2&=&2\cosh ebs \frac{\cos \nu eas -\cos eas}{\sin^2 eas},
  \nonumber\\  
N_3&=&-\frac{1-\cos eas\, \cos \nu eas}{\sin eas}\frac{\cosh \nu ebs\,
  \cosh ebs -1}{\sinh ebs} +\sin \nu eas\, \sinh \nu ebs. \label{111225}
\end{eqnarray}
The scalars $v^2_{\|,\bot}$ are given by certain combinations of the 
invariants and can be found in Eq. \re{111219}. The contact term given
in Eq. \re{111210} contributes equally to the $\Pi_i$'s,
\begin{equation}
\text{c.t.}=-\E^{-\I m^2 s} k^2(1-\nu^2), \label{111226}
\end{equation}
but does not modify the function $\Theta$, which is already finite. 

Note that Eq. \re{111223} almost appears in a diagonalized form except
for the term $\Theta\, Q^{\mu\nu}$. While $P_0^{\mu\nu}$ indeed
projects onto an eigenspace of $\Pi^{\mu\nu}$ with eigenvalue $\Pi_0$,
this is generally not the case for the projectors
$P_{\|,\bot}^{\mu\nu}$, due to $\Theta\, Q^{\mu\nu}$. Although a
further diagonalization is straightforward, we will not bother to
write it down, since we only need the trace of $\Pi^{\mu\nu}$, which
is simply given by:
\begin{equation}
\Pi^\mu{}_\mu=\Pi_0+\Pi_\|+\Pi_\bot, \qquad Q^\mu{}_\mu=0. \label{7}
\end{equation}
In the actual two-loop calculation, the contact terms can be omitted
for two reasons: first, it does not contribute to the thermal part,
since the latter is finite; secondly, for the zero-temperature
Lagrangian, a renormalization procedure is required anyway and, in
particular, the mass renormalization would not be covered by an
inclusion of the contact terms. 

Inserting Eq. \re{111224} into Eq. \re{7} brings us to the explicit
representation of the trace:
\begin{equation}
\Pi^\mu{}_\mu\!=\frac{\alpha}{2\pi}\!
\int\limits_0^\infty\!\frac{ds}{s}\!  
\!\int\limits_{-1}^1\!\frac{d\nu}{2} \frac{\E^{-\I
      s\phi_0}}{a^2\!+\!b^2} \frac{eas\,ebs}{\sin eas \sinh ebs} 
\Bigl[z_k (\tilde{N}_2 -\tilde{N}_1) +k^2\bigl( 2N_0(a^2\!+\!b^2)
  +b^2\tilde{N}_2 +a^2\tilde{N}_1\bigr)\!\Bigr]\!. \label{10}
\end{equation}
This is the desired expression which is required in Eq. \re{5}. For
reasons of convenience, it is useful to rewrite the function $\phi_0$
in terms of the variables $k^2$ and $z_k$. For this, we insert
Eq. \re{111219} into the first line of Eq. \re{111225}; a
reorganization yields:
\begin{equation}
\E^{-\I s\phi_0}=\E^{-\I m^2s}\, \E^{-A_z \, z_k}\, \E^{-A_k\, k^2},
\label{15}
\end{equation}
where we implicitly defined:
\begin{eqnarray}
A_z&:=& \frac{\I s}{2}\, \frac{1}{a^2+b^2}\, 
  \left( \frac{\cos \nu eas -\cos eas}{eas \sin eas} + \frac{\cosh \nu
  ebs -\cosh ebs}{ebs \sinh ebs} \right), \nonumber\\
A_k&:=& \frac{\I s}{2}\, \frac{1}{a^2+b^2}\, 
  \left( b^2\frac{\cos \nu eas -\cos eas}{eas \sin eas} -a^2
  \frac{\cosh \nu  ebs -\cosh ebs}{ebs \sinh ebs}
  \right). \label{16}\\ 
\end{eqnarray}
This provides us with the required necessities for the two-loop
calculation in Sec. 2.

\section{Finite-Temperature Coordinate Frame}

In order to make the paper self-contained, we briefly review the
construction of the finite-temperature coordinate frame as introduced
in \cite{gies99a}, and then apply it to the present problem.

First, we define the {\em vierbein} $e^{A\mu}$ which mediates between
the given system labelled by $\mu,\nu,\dots=0,1,2,3$ and the desired
system labelled by the (Lorentz) indices $A,B,\dots=0,1,2,3$ by:
\begin{eqnarray}
e_0{}^\mu&:=& u^\mu,\nonumber\\
e_1{}^\mu&:=& \frac{u_\alpha F^{\alpha\mu}}{\sqrt{{\cal E}}},
\nonumber\\ 
e_2{}^\mu&:=& \frac{1}{\sqrt{d}} \bigl( u^\alpha F_{\alpha\beta}
  F^{\beta\mu} -{\cal E}\, e_0{}^{\mu} \bigr), \nonumber\\
e_3{}^\mu&:=&\epsilon^{\alpha\beta\gamma\mu}\, e_{0\alpha}\,
e_{1\beta}\, e_{2\gamma}, \label{2.3}
\end{eqnarray}
where the quantity $d$ abbreviates the combination of invariants:
\begin{equation}
d:=2{\cal F}{\cal E}-{\cal G}^2+{\cal E}^2. \label{2.4}
\end{equation}
The vierbein satisfies the identity
\begin{equation}
e_{A\mu}\, e_B{}^\mu =g_{AB}\equiv\text{diag}(-1,1,1,1), \label{2.5}
\end{equation}
where $g_{AB}\sim g^{AB}$ denotes the metric which raises and lowers
capital indices. By a direct computation, we can transform the field
strength tensors and the heat-bath vector:
\begin{eqnarray}
n^A&:=&g^{AB}e_{B}{}^\mu\,n_\mu=(T,0,0,0), \nonumber\\
F_{AB}&:=&e_{A\mu}F^{\mu\nu}e_{B\nu}
  =\left(\begin{array}{cccc}
    0      &\sqrt{{\cal E}}   &      0     &      0    \\
  -\sqrt{{\cal E}}&      0    &\sqrt{d/{\cal E}}  &      0    \\
    0      &-\sqrt{d/{\cal E}}&  0 &-{\cal G}/\sqrt{{\cal E}}\\
    0      &      0    &{\cal G}/\sqrt{{\cal E}}  &      0    
   \end{array}\right). \label{2.8}
\end{eqnarray}
Obviously, the new system corresponds to the heat-bath rest frame with
the spatial axes oriented along the electromagnetic field in some
sense. The components of the field strength tensor are now given by
combinations of the invariants. 

In order to determine the form of $z_k=k_\mu F^{\mu\alpha} k_\nu
F^\nu{}_\alpha\equiv -k^A F_{AC} F^C{}_B k^B$, we need the square of
the field strength tensor:
\begin{equation}
F^2_{AB}\equiv  F_{AC} F^C{}_B =
\left( \begin{array}{cccc}
 -{\cal E}   &       0        &     \sqrt{d}     &        0\\
    0        &{\cal E}-\frac{d}{{\cal E}}&0      &-\frac{\sqrt{d}{\cal
                                                  G}}{{\cal E}}\\
\sqrt{d}     &       0        &-\frac{{\cal G}^2+d}{{\cal E}}&0\\
0&-\frac{\sqrt{d}{\cal G}}{{\cal E}}&0&-\frac{{\cal G}^2}{{\cal E}}
\end{array}\right). \label{20}
\end{equation}
This allows us to write $z_k$ in the form:
\begin{equation}
z_k={\cal E}\, (k^0)^2 -2\sqrt{d}\, k^0k^2 +(2{\cal F}+{\cal E})\,
(k^2)^2 +\left(\!\frac{d}{{\cal E}}-{\cal E}\!\right) (k^1)^2
+2\frac{\sqrt{d}{\cal G}}{{\cal E}}\, k^1k^3 +\frac{{\cal G}^2}{{\cal
    E}}\, (k^3)^2, \label{21a}
\end{equation}
where $k^0,k^1,k^2,k^3$ represent the components of the rotated
momentum vector $k^A=e^A{}_\mu k^\mu$.

Now we can finally determine the desired form for the exponent in
Eq. \re{18} in terms of finite-temperature coordinates:
\begin{eqnarray}
A_z z_k +A_k k^2 \!\!&=&\!\! \bigl( A_k +(a^2\!-\!b^2\!+\!{\cal E})
A_z\bigr)\! \left(\! k^2\!-\!{\scriptstyle
    \frac{A_z\sqrt{d}}{A_z(2{\cal  F} +{\cal  E}) +A_k}}\, k^0\! 
\right)^2 -\frac{(A_k\!+\!a^2A_z)(A_k\!-\!b^2A_z)}{ A_k
  +(a^2\!-\!b^2\!+\!{\cal  E}) A_z} \, (k^0)^2 \nonumber\\
&&\!\! +\left(\! A_z\frac{a^2b^2}{{\cal E}} +A_k\!\right)\!\left(\!k^3
   +{\scriptstyle \frac{A_z \frac{\sqrt{d}{\cal G}}{{\cal E}}}{A_z
       \frac{{\cal G}^2}{{\cal E}} +A_k}}\, k^1\!\right)
 +\frac{(A_k+a^2A_z)(A_k-b^2A_z)}{ A_k\frac{a^2b^2}{{\cal E}} + A_k}\,
 (k^1)^2, \nonumber\\
&&\label{23}
\end{eqnarray}
where again, $k^0,k^1,k^2,k^3$ represent the components of $k^A$. 

\section{Two-loop Effective Action of QED at Zero Temperature} 

Dedicating this appendix to the derivation of the zero-temperature
two-loop Lagrangian has two reasons: first, we want to make
contact to well-known results, which serve as a check for our
computations; secondly, our results will represent a generalization of
the work of Dittrich and Reuter \cite{ditt85}, who considered purely
magnetic fields, to the case of constant electromagnetic
fields. 

Here, we will only give a version of the unrenormalized effective
Lagrangian, since the renormalization program requires detailed
investigations which are beyond the scope of the present
work. In particular, the mass renormalization has to be treated with
great care \cite{flie97}, taking a correct matching of the imaginary parts
of the effective action into account. 

In Eq. \re{29}, we achieved a separation of the thermal and the
zero-temperature parts in the integral $I_1$; concentrating on the
zero-$T$ case, we found:
\begin{equation}
I_1^{T=0}=\frac{1}{16\pi^2} \E^{-\I m^2 s}\,
\frac{1}{q_a\,q_b}, \label{A.1}
\end{equation}
where $q_a=(A_k+a^2A_z)$ and $q_b=(A_k-b^2A_z)$, and $A_k$ and $A_z$
are defined in Eq. \re{16}. We also need the second integral
$I_2^{T=0}$, which is related to $I_1^{T=0}$ by Eq. \re{17}, leading
us to:
\begin{equation}
I_2^{T=0}=\frac{\E^{-\I m^2s}}{16\pi^2} \int\limits_0^\infty ds'
\left[ 
\frac{a^2}{(s'+q_a)^2(s'+q_b)} 
-\frac{b^2}{(s'+q_a)(s'+q_b)^2}
\right].\label{A.2}
\end{equation}
The integration over $s'$ can be carried out with elementary
techniques:
\begin{equation}
I_2^{T=0}=\frac{\E^{-\I m^2s}}{16\pi^2}\left[ a^2 \left
    ( \frac{1}{q_a(q_b-q_a)} -\frac{1}{(q_b-q_a)^2} \ln
    \frac{q_b}{q_a}\right) -b^2\left( \frac{1}{q_b(q_a-q_b)}
    +\frac{1}{(q_b-q_a)^2} \ln
    \frac{q_b}{q_a}\right)\right]\!. \label{A.3}
\end{equation}
Inserting these $T=0$-contributions \re{A.1} and \re{A.3} into Eq.
\re{11a} and reorganizing the result a bit, we finally arrive at:
\begin{eqnarray}
{\cal L}^2&=&-\frac{\alpha}{(4\pi)^3} \int\limits_0^\infty \frac{ds}{s}
\int\limits_{-1}^1 \frac{d\nu}{2}\,\E^{-\I m^2 s}\,  \frac{eas\,
ebs}{\sin eas \sinh ebs} \label{A.7}\\
&&\qquad\quad \times \left[ \frac{2N_0+\tilde{N}_2}{q_a(q_b-q_a)} +
  \frac{2N_0 +\tilde{N}_1}{q_b(q_a-q_b)}
  -\frac{\tilde{N}_2-\tilde{N}_1}{(q_b-q_a)^2} \ln \frac{q_b}{q_a}
\right], \nonumber
\end{eqnarray}
where $N_0$, $\tilde{N}_1$, and $\tilde{N}_2$ are functions of the
integration variables and the invariants $a$ and $b$, and are defined
in Eq. \re{111225}, whereas $q_a$ and $q_b$ after insertion of Eq.
\re{16} can be written as:
\begin{equation}
q_a=\frac{\I s}{2} \frac{\cos \nu eas -\cos eas}{eas \sin eas},\qquad 
q_b=\frac{\I s}{2} \frac{\cosh ebs -\cosh \nu ebs}{ebs \sinh
  ebs}. \label{A.4}
\end{equation}
Equation \re{A.7} represents our final result for the unrenormalized
two-loop effective Lagrangian of QED for arbitrary constant
electromagnetic fields. In the limit of vanishing electric fields,
we recover exactly the findings of \cite{ditt85}; hence our comparably
compact representation generalizes the results of \cite{ditt85} to
arbitrary constant electromagnetic fields. 

\section*{Acknowledgments}

I would like to thank Professor W. Dittrich for helpful
discussions and for carefully reading the manuscript. I am also
grateful to Dr R. Shaisultanov for valuable comments and especially
for drawing my attention to photon-photon scattering.

\end{document}